\documentclass[10pt,letterpaper]{article}
\usepackage[top=0.85in,left=1in,footskip=0.75in]{geometry}

\usepackage{amsmath,amssymb}

\usepackage[utf8x]{inputenc}
\usepackage{textcomp,marvosym}
\usepackage{cite}
\usepackage{nameref}
\usepackage{hyperref}
\usepackage[table,dvipsnames]{xcolor}
\usepackage{array}

\newcolumntype{+}{!{\vrule width 2pt}}
\newlength\savedwidth

\usepackage{setspace} 

\raggedright
\usepackage[aboveskip=1pt,labelfont=bf,labelsep=period,justification=raggedright,singlelinecheck=off]{caption}

\usepackage{xcolor}
\definecolor{MyRed}{rgb}{0.51,0.07,0.15}
\definecolor{MyBlue}{rgb}{0.16,0.34,0.58}
\newcommand\changed[1]{#1} 

\usepackage{multirow,subfigure}
\usepackage{pdflscape}

\date{}

\usepackage{lastpage,fancyhdr,graphicx}
\begin{document}
\vspace*{0.2in}

\begin{flushleft}
{\Large
\textbf\newline{Gaussian graphical models reveal inter-modal and inter-regional conditional dependencies of brain alterations in Alzheimer's disease}
}
\newline
\\
Martin Dyrba\textsuperscript{1*}, 
Reza Mohammadi\textsuperscript{2**}, 
Michel J. Grothe\textsuperscript{1},
Thomas Kirste\textsuperscript{3},
Stefan J. Teipel\textsuperscript{1,4},
for the Alzheimer's Disease Neuroimaging Initiative\textsuperscript{\textpilcrow}
\\
\bigskip
\textbf{1} German Center for Neurodegenerative Diseases (DZNE), Rostock/Greifswald, Rostock, Germany
\\
\textbf{2} Department of Operation Management, Amsterdam Business School, University of Amsterdam, Amsterdam, The Netherlands
\\
\textbf{3} Mobile Multimedia Information Systems Group (MMIS), University of Rostock, Rostock, Germany
\\
\textbf{4} Clinic for Psychosomatics and Psychotherapeutic Medicine, University Medicine Rostock, Rostock, Germany
\\
\bigskip

%
%




\textpilcrow Data used in preparation of this article were obtained from the Alzheimer's Disease Neuroimaging Initiative (ADNI) database (\url{http://adni.loni.usc.edu}). As such, the investigators within the ADNI contributed to the design and implementation of ADNI and provided data but did not participate in analysis or in the writing of this report. A complete listing of ADNI investigators can be found at: \url{http://adni.loni.usc.edu/wp-content/uploads/how_to_apply/ADNI_Acknowledgement_List.pdf}.
\bigskip

* \textbf{Corresponding author at:} German Center for Neurodegenerative Diseases (DZNE), c/o Zentrum für Nervenheilkunde, Gehlsheimer Straße 20, D-18147 Rostock, Germany, phone +49-381-494-9482, fax +49-381-494-9472, email martin.dyrba@dzne.de
\\
** \textbf{Corresponding author at:} Department of Operation Management, Amsterdam Business School, P.O. Box 15953, 1001 NL Amsterdam, The Netherlands, email a.mohammadi@uva.nl
\bigskip

\textbf{Short title:} Conditional dependency networks in AD.
\\
\textbf{Keywords:} Alzheimer's disease,
	mild cognitive impairment,
	conditional dependency networks,
	Gaussian graphical models,
	graph-theoretical analysis,
	small-world network.

\end{flushleft}


\section*{Abstract}

Alzheimer's disease (AD) is characterized by a sequence of pathological changes, which are commonly assessed in vivo using various brain imaging modalities such as magnetic resonance imaging (MRI) and positron emission tomography (PET).
Currently, the most approaches to analyze statistical associations between regions and imaging modalities rely on Pearson correlation or linear regression models.
However, these models are prone to spurious correlations arising from uninformative shared variance and multicollinearity.
Notably, there are no appropriate multivariate statistical models available that can easily integrate dozens of multicollinear variables derived from such data, being able to utilize the additional information provided from the combination of data sources.
Gaussian graphical models (GGMs) can estimate the conditional dependency from given data, which is conceptually expected to closely reflect the underlying causal relationships between various variables.
Hence, we applied GGMs to assess multimodal regional brain alterations in AD.

We obtained data from N=972 subjects from the Alzheimer’s Disease Neuroimaging Initiative. The mean amyloid load (AV45-PET), glucose metabolism (FDG-PET), and gray matter volume (MRI) were calculated for each of the 108 cortical and subcortical brain regions.
GGMs were estimated using a Bayesian framework for the combined multimodal data and the resulted conditional dependency networks were compared to classical covariance networks based on Pearson correlation.
Additionally, graph-theoretical network statistics were calculated to determine network alterations associated with disease status.

The resulting conditional dependency matrices were much sparser ($\approx10\%$ density) than Pearson correlation matrices ($\approx50\%$ density).
Within imaging modalities, conditional dependency networks yielded clusters connecting anatomically adjacent regions.
For the associations between different modalities, only few region-specific connections were detected.
Network measures such as small-world coefficient were significantly altered across diagnostic groups, with a biphasic u-shape trajectory, i.e. increased small-world coefficient in early mild cognitive impairment (MCI), similar values in late MCI, and decreased values in AD dementia patients compared to cognitively normal controls.

In conclusion, GGMs removed commonly shared variance among multimodal measures of regional brain alterations in MCI and AD, and yielded sparser matrices compared to correlation networks based on the Pearson coefficient.
Therefore, GGMs may be used as alternative to thresholding-approaches typically applied to correlation networks to obtain the most informative relations between variables.

\section{Introduction}

Alzheimer's disease (AD) is characterized by a range of pathological brain alterations that can be assessed in vivo using various neuroimaging methods, including MRI and PET. 
Several studies suggest that information obtained from combining different imaging modalities could provide reliable markers of cerebral reserve capacity and might be used to predict and monitor the evolution of AD and its relative impact on cognitive domains in pre-clinical, prodromal, and dementia stages of AD (see e.g. reviews \cite{Teipel.2015,Teipel.2016b}).
\changed{However, there is still an unmet need for appropriate analysis methods for assessing statistical associations between individual brain regions and between different pathology markers derived from multiple neuroimaging modalities.}
Up to date, multimodal studies employ one of the following approaches:
	(i) Correlation of pathology maps on a voxel level \cite{Altmann.2015,Grothe.2016b,LaJoie.2012};
	(ii) linear regression analysis with a-priori specified regions-of-interest \cite{Buckner.2005, Seeley.2009, Villain.2010, Kljajevic.2014, Chang.2015, Grothe.2016a, Teipel.2016};
	(iii) stratification of subjects into distinct groups (e.g. amyloid positive/negative) to compare differences in other imaging modalities \cite{Buckner.2005, Kljajevic.2014, Grothe.2016a};
	(iv) comparison of graph-theoretical measures and statistics between modalities \cite{Stam.2006, Buckner.2009, Zhou.2012, Sepulcre.2013, Sepulcre.2017}; and
	(v) estimation of generative models for comparing spreading mechanisms of amyloid-$\beta$ deposition and its contribution to neurodegeneration \cite{IturriaMedina.2017,Dyrba.2017,Torok.2018}. 
%
Commonly employed statistical models, such as linear regression analysis, provide limited ability to assess the interactions between dozens of variables in the same model, as they cannot derive reliable estimates regarding the individual contribution of highly collinear predictors and suffer from variance inflation \cite{Dormann.2013}.
Calculation of covariance/connectivity matrices based on the Pearson correlation between each pair of variables has led to practical problems in deriving meaningful results, i.e. these matrices are commonly thresholded to an a-priori defined density and binarized \cite{Buckner.2009,Zhou.2012,Sepulcre.2013}. More recently, summary statistics based on graph-theory have been proposed \cite{Watts.1998,Stam.2006} and are currently widely applied \cite{Buckner.2009,Zhou.2012,Sepulcre.2013,Sepulcre.2017}. However, this approach has 
been criticized, as for instance, group differences in small-worldness of the brain network might be sensitive to the specific density threshold \cite{Hlinka.2017,Martensson.2018}.

\begin{figure*}[!ht]
	\centering
		\subfigure{\includegraphics[width=0.25\textwidth]{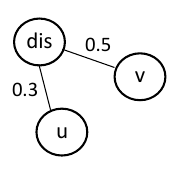}}
			\llap{\parbox[b]{1.7in}{\textbf{(A)}\\\rule{0ex}{1.4in}}}\hfill
		\subfigure{\includegraphics[width=0.3\textwidth]{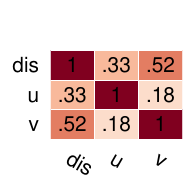}}
			\llap{\parbox[b]{2.1in}{\textbf{(B)}\\\rule{0ex}{1.4in}}}\hfill
		\subfigure{\includegraphics[width=0.3\textwidth]{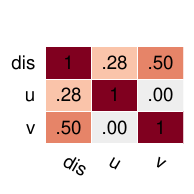}}
			\llap{\parbox[b]{2.1in}{\textbf{(C)}\\\rule{0ex}{1.4in}}}
	\caption{Simple example for spurious correlations.
	\textbf{(A)} True dependency graph. The node $u$ is statistically independent from $v$ given the node $dis$, formally $p(u,v|dis) = p(u|dis)p(v|dis)$.
	\textbf{(B)} Pearson correlation matrix, showing a "spurious" correlation between nodes u and v. 
		Notably, when considering only $u$ and $v$ alone, the independence assumption does not hold; formally $p(u,v) \neq p(u)p(v)$.
	\textbf{(C)} Partial correlation matrix derived from Gaussian graphical models. 
		Using this model, we can approximately recover the underlying dependency structure, with $u \perp v | dis \implies cor(u,v|dis)=0$.}
	\label{fig:example}
\end{figure*}

We suggest the application of Gaussian graphical models (GGMs), which are able to estimate the \textit{partial} correlation between various multicollinear predictors \cite[ch. 7.3]{Hastie.2013}.
GGMs yield sparse conditional dependency matrices, that are conceptually expected to closer reflect the underlying causal relationships \cite[chapter~21.7]{Bontempi.2015,Koller.2009}. This makes GGMs an interesting candidate for studying properties of the brain network;
an example is illustrated in Figure \ref{fig:example}.
The partial correlation derived from GGMs is conceptually similar to the partial correlation obtained from a series of linear regression models, which estimate the statistical association of the dependent and independent variables while controlling for the confounding variables.
Additionally, GGMs extend this concept by estimating the partial correlation matrix as a set of coupled regression problems, in contrast to separate regression problems modeled by traditional linear regression \cite[ch. 7.3]{Meinshausen.2006,Hastie.2013}.
Technically, GGMs are naively realized by matrix inversion of the covariance matrix.
In more robust and efficient approaches, regularization techniques \cite{Meinshausen.2006,Ravikumar.2011,Ryali.2012,Cai.2013,Wang.2016} or efficient sampling schemes \cite{Mohammadi.2015,Mohammadi.2019} are applied.

\changed{In this paper, we tested the applicability and clinical utility of GGMs to reveal the conditional dependency structure between regional pathology measures.}
For this purpose, we assessed inter-regional statistical associations within and between three main imaging markers of Alzheimer's disease using GGMs based on a whole-cortex parcellation of the brain. The assessed imaging markers included amyloid-$\beta$ deposition (florbetapir/AV45-PET), glucose metabolism (FDG-PET), and gray matter volume ($T_1$-weighted MRI).
Based on our previous results with only six representative brain regions \cite{Dyrba.2017}, we hypothesized that regional amyloid deposition has low contribution to gray matter atrophy, whereas hypometabolism was expected to be stronger related to atrophy.
Further, we expected a few hub-nodes influencing pathology in other regions.
For graph-theoretical measures, we expected a linear trajectory of decreasing clustering coefficient and increasing path length with stronger disease severity, as previously reported in the literature for connectivity analyses based on Pearson correlation \cite{He.2008,Yao.2010,Li.2012,Morbelli.2012,Tijms.2013,Pereira.2016,John.2017,Titov.2017}.

\section{Material and Methods}

\subsection{Study participants}

Data were obtained from the Alzheimer's Disease Neuroimaging Initiative (ADNI) database (\url{http://adni.loni.usc.edu}). The ADNI was launched in 2003 by the National Institute on Aging, the National Institute of Biomedical Imaging and Bioengineering, the Food and Drug Administration, private pharmaceutical companies, and non-profit organizations, with the primary goal of testing whether neuroimaging, neuropsychological, and other biological measurements can be used as reliable in vivo markers of Alzheimer's disease pathogenesis. A complete description of ADNI and up-to-date information is available at \url{http://www.adni-info.org}.
For this study, 529 subjects with amnestic mild cognitive impairment (MCI), 189 patients with Alzheimer's dementia (AD), and 254 cognitively healthy control subjects (CN) were selected from the ADNI-GO, ADNI-2 and ADNI-3 extensions of the ADNI project, based on the availability of concurrent structural MRI, FDG-PET, amyloid-sensitive AV45-PET, and neuropsychological assessments. In ADNI, two MCI subgroups exist, which only differ by the less severe impairment of memory function for \textit{early MCI} (EMCI) compared to \textit{late MCI} (LMCI) subjects.
Detailed inclusion criteria for the diagnostic categories can be found at the ADNI website (\url{http://adni.loni.usc.edu/methods}, ADNI2 manual page 27).
Demographics and neuropsychological profiles of the different diagnostic groups are summarized in Table~\ref{tab:sample}.

\subsection{Imaging data and feature extraction}

ADNI-GO/2 MRI, FDG- and AV45-PET data were downloaded from the ADNI image archive.
\changed{ADNI-GO/2 MRI data were acquired on multiple 3T MRI scanners using scanner-specific T1-weighted sagittal 3D MP-RAGE/IR-SPGR sequences. To increase signal uniformity across the multicenter scanner platforms, original T1 acquisitions underwent standardized image preprocessing correction steps (\url{http://adni.loni.usc.edu/methods/mri-tool/mri-pre-processing/}). 
FDG- and AV45-PET data were acquired on multiple instruments of varying resolution and following different platform-specific acquisition protocols. Similar to the MRI data, PET data in ADNI were also subject to standardized image preprocessing correction steps, with the aim of increasing data uniformity across the multicenter acquisitions (\url{http://adni.loni.usc.edu/methods/pet-analysis-method/pet-analysis/}).}
Imaging data were processed by using statistical parametric mapping (SPM8, Wellcome Centre for Human Neuroimaging, University College London) and the VBM8 toolbox (Structural Brain Mapping Group, University of Jena) implemented in MATLAB R2013b (Math-Works, Natick, MA) as previously described in \cite{Grothe.2016a,Grothe.2016b}.
First, MRI T1 scans were segmented into gray matter, white matter and cerebrospinal fluid partitions using the segmentation routine of the VBM8 toolbox.
Then, the resulting gray matter and white matter segments were spatially normalized to an aging/AD-specific reference template \cite{Grothe.2013} using the DARTEL algorithm.
Additionally, voxel values of the normalized gray matter segments were modulated for volumetric changes introduced by the high-dimensional normalization, such that the total amount of gray matter volume present before warping was preserved.
Each subject's FDG- and AV45-PET scans were rigidly coregistered to the corresponding skull-stripped T1 scan.
Then, the PET scans were corrected for partial volume effects using a three-compartment model and the MRI-derived tissue segments \cite{MullerGartner.1992,GonzalezEscamilla.2017}.
Corrected PET scans were spatially normalized (without modulation) by applying the deformation fields of the T1-weighted scans.
\changed{All original data and normalized scans were visually inspected to ensure a high quality of the data.}
Subsequently, mean gray matter volumes and mean FDG-/AV45-PET uptake values were calculated for 108 cortical and subcortical regions defined by the Harvard-Oxford atlas \cite{Desikan.2006} after projecting the atlas to the aging/AD-specific reference space and removing voxels with a gray matter probability of less than 50\% in the aging/AD template.
Finally, regional gray matter volumes were proportionally scaled by total intracranial volume (TIV), regional FDG-PET values were proportionally scaled to pons uptake, and regional AV45-PET values were proportionally scaled to whole-cerebellum uptake.
To be able to directly compare the different modalities, all regional values were normalized using the congitively normal subjects as reference group \cite{LaJoie.2012}.
As described previously \cite{Dyrba.2017}, we used the so-called $W$-scores, which are analogous to $Z$-scores but are adjusted for specific covariates; age, gender, and education in the present case.
Like $Z$-scores, $W$-scores have a mean value of 0 and a standard deviation of 1 in the control group, and values of $+1.65$ and $-1.65$ correspond to the $95^{th}$ and $5^{th}$ percentiles, respectively.
To calculate the $W$-scores, regression models were estimated for the control group using age, gender, and education as independent variables and the mean value of each region as dependent variable.
Then, $W$-scores were computed using $W=(x_{ij}-e_{ij})/s_{res,j}$; with $x_{ij}$ being the $i^{th}$ subject's raw value for region $j$; $e_{ij}$ being the value expected for region $j$ in the control group for the $i^{th}$ subject's age, gender, and education; and $s_{res,j}$ being the standard deviation of the residuals for region $j$ in controls.

\subsection{Statistical modeling}

Graphical models provide an effective way for describing statistical patterns in multivariate data and for estimating the conditional dependency between the various brain regions and imaging modalities based on GGMs \cite{Lauritzen.1996,Mohammadi.2015}.
For data following a multivariate normal distribution, undirected GGMs are commonly used.
In these graphical models, the graph structure is directly characterized by the precision matrix, i.e. the inverse of the covariance matrix: non-zero entries in the precision matrix show the edges in the conditional dependency graph.
Notably, simple inversion of the covariance matrix usually does not work in real world data sets, as already slight noise in the empirical data causes the precision matrix to contain almost no zero entries.
To overcome this problem, regularization techniques or efficient sampling algorithms have been proposed that reduce the effect of noise by additionally employing a sparsity assumption and, thus, only detect the most probable conditional dependencies.
For our analyses, we employed a computationally efficient Bayesian framework implemented in the R package BDgraph. More specifically, this framework implements a continuous-time birth-death Markov process for estimating the most probable graph structure and edge weights that correspond to the observed partial correlations \cite{Mohammadi.2015,Mohammadi.2019}.
For this study, BDgraph was substantially extended by multi-threaded parallel processing and marginal pseudo-likelihood approximation to speed up computations.

\subsection{Experimental setup}

First, we estimated GGMs based on the combined data of EMCI, LMCI and AD patients to study the conditional dependency between brain regions and modalities. Second, we estimated GGMs for each diagnostic group separately to assess alterations of the graph structures.
For the combined model, regional $W$-scores of all MCI and AD patients ($N=718$) and all three imaging modalities were entered. 
Initially, we took all $108$ cortical and subcortical regions included in the Harvard-Oxford atlas \cite{Desikan.2006} into consideration, corresponding to $P=3*108=324$ variables. The sampling process included 1,000,000 burn-in iterations\footnote{For Markov chain Monte Carlo (MCMC) methods, burn-in refers to the practice of discarding an initial portion of the Markov chain sample, so that the chain can reach a stationary distribution. Thus, the effect of randomly chosen initial values on the posterior inference is minimized.},
starting from a random estimate for the inverse covariance matrix and converging to estimates with higher posterior probability giving the training data. The burn-in iterations were then discarded, and subsequently 150,000 sampling iterations followed to obtain the estimates for the inverse covariance matrix.
Because results were showing a strong left--right hemisphere symmetry, we repeated model estimation including only the 54 regions in the left hemisphere ($P=3*54=162$ variables) to increase model stability.
From the final model we set a probability threshold of $P_{avg}>0.5$ for selecting the edges, with the notion that a specific edge was considered to be present if it existed in at least half of all model iterations \cite{Madigan.1996}.
For the second analysis of group differences, we estimated individual GGMs for each group based on the multimodal data of the left hemisphere. Sampling was again performed with 1,000,000 burn-in iterations followed by 150,000 sampling iterations.

\changed{For comparison, these analyses were also repeated (i) using data of the right hemisphere to validate the results} and (ii) using the traditional approach of constructing correlation networks based on the Pearson correlation coefficient.



\subsection{Graph-theoretical analyses}

To assess group differences of the estimated graph structure we calculated the three graph-theoretical measures that are most commonly reported in the literature; clustering coefficient, characteristic path length, and their ratio, the small-world coefficient.
The path length quantifies the distance of connections between two nodes along the shortest path. The weighted characteristic path length is the average minimum distance between a node $i \in N$ and all other nodes, $L_i=\sum_{j\in N,j\neq i}d_{ij}/(n-1)$, where $d_{ij}=\sum_{a_{uv}\in g_{i \leftrightarrow j}}\omega_{uv}$ is the shortest weighted path length between $i$ and $j$, $g_{i \leftrightarrow j}$ defines the shortest path, and $\omega_{uv}$ defines the distance between two nodes. Here, the distance matrix was defined as $\Omega=1-abs(\Theta)$, that is one minus the absolute pair-wise partial correlation as derived from the GGMs or the absolute Pearson coefficient, respectively \cite{Rubinov.2010}.
The weighted clustering coefficient indicates the inter-connectedness of neighboring nodes $C_i=2t_i/(k_i(k_i-1))$, where $t_i=0.5\sum_{j,h\in N}(\omega_{ij}\omega_{ih}\omega_{jh})^{1/3}$ is the geometric mean of triangles around node $i$, and where $k_i=\sum_{j\in N}a_{ij}$ is the number of nodes connected to node $i$ \cite{Onnela.2005,Rubinov.2010}. $k_i$ is often referred to as the \textit{degree} of the node $i$, and the link status $a_{ij}=1$ if node $i$ is connected to another node $j$, or $a_{ij}=0$ otherwise.
The small-world coefficient is defined as the ratio of the clustering coefficient $C$ and characteristic path length $L$ in comparison to a random network, $S=(C/C_{rand})/(L/L_{rand})$, with $S\gg1$ in small-world networks \cite{Rubinov.2010}. To simplify calculations, we omitted defining a random network to estimate $C_{rand}$ and $L_{rand}$, and directly took the ratio $S_i=C_i/L_i$ for group comparisons.
Notably, we later report the distribution of graph measures for single regions, as the dependency measures were derive from the whole group of subjects.
\changed{Graph metrics were compared between diagnostic groups using analysis of variance (ANOVA) and Tukey's honest significant difference tests.}

\section{Results}

\subsection{Conditional dependency of Alzheimer's pathology}

\begin{figure*}[!ht]
	\centering
		\subfigure{\includegraphics[width=0.5\textwidth]{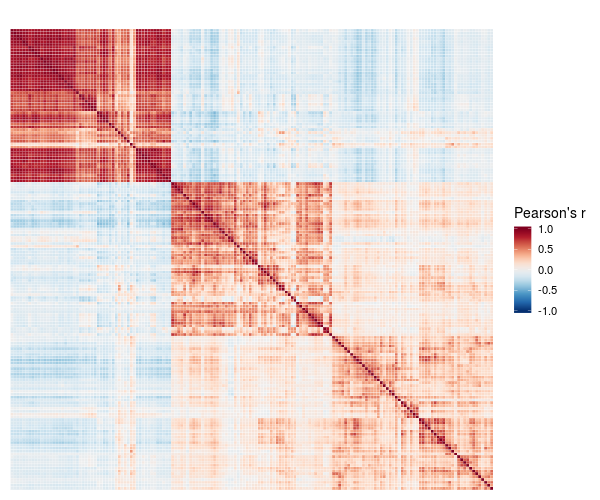}}\hfill
		\subfigure{\includegraphics[width=0.5\textwidth]{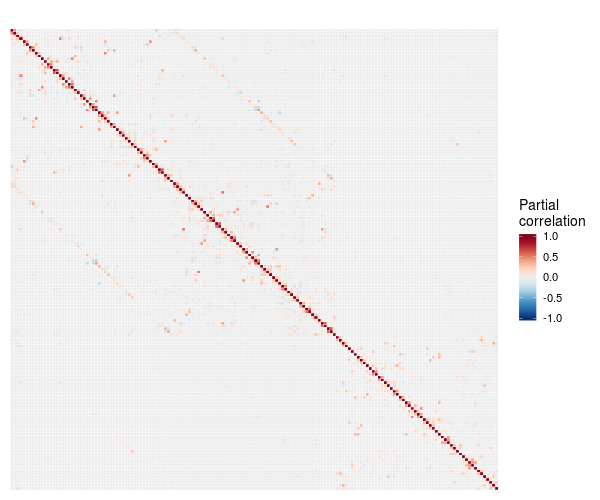}}
	\caption{Pearson correlation matrix (left) and partial correlation matrix (right) for the three imaging modalities (left hemisphere data only) estimated for the combined data of EMCI, LMCI and AD patients.
	For better readability, each individual block of the partial correlation matrix is shown in Figures~\ref{fig:parcor-mat-lh-amy-amy}--\ref{fig:parcor-mat-lh-vol-vol} and Supplementary Figures \ref{sfig:parcor-mat-lh-amy-fdg}--\ref{sfig:parcor-mat-lh-amy-vol}.\\
	EMCI/LMCI: early and late amnestic mild cognitive impairment, AD: Alzheimer's dementia.}
	\label{fig:corr-mat-whole-brain}
\end{figure*}

The conditional dependency matrix obtained using the GGM approach for all region of the left hemisphere is given in Figure~\ref{fig:corr-mat-whole-brain} (right).
For the partial correlation between all pairs of brain regions, we obtained 960 significant associations ($7\%$ network density) surviving the posterior probability threshold of $P>0.5$ \changed{(see Supplementary Figure \ref{sfig:plinks-lh} showing the probability of links).}
For comparison, the Pearson correlation matrix is given in Figure~\ref{fig:corr-mat-whole-brain} left).
We obtained approximately 6,000 significant Pearson correlations ($P<0.05$, Bonferroni corrected), corresponding to a network density of $46\%$ of the total number of possible edges. 

\begin{figure*}[!ht]
	\centering
		\includegraphics[width=\textwidth]{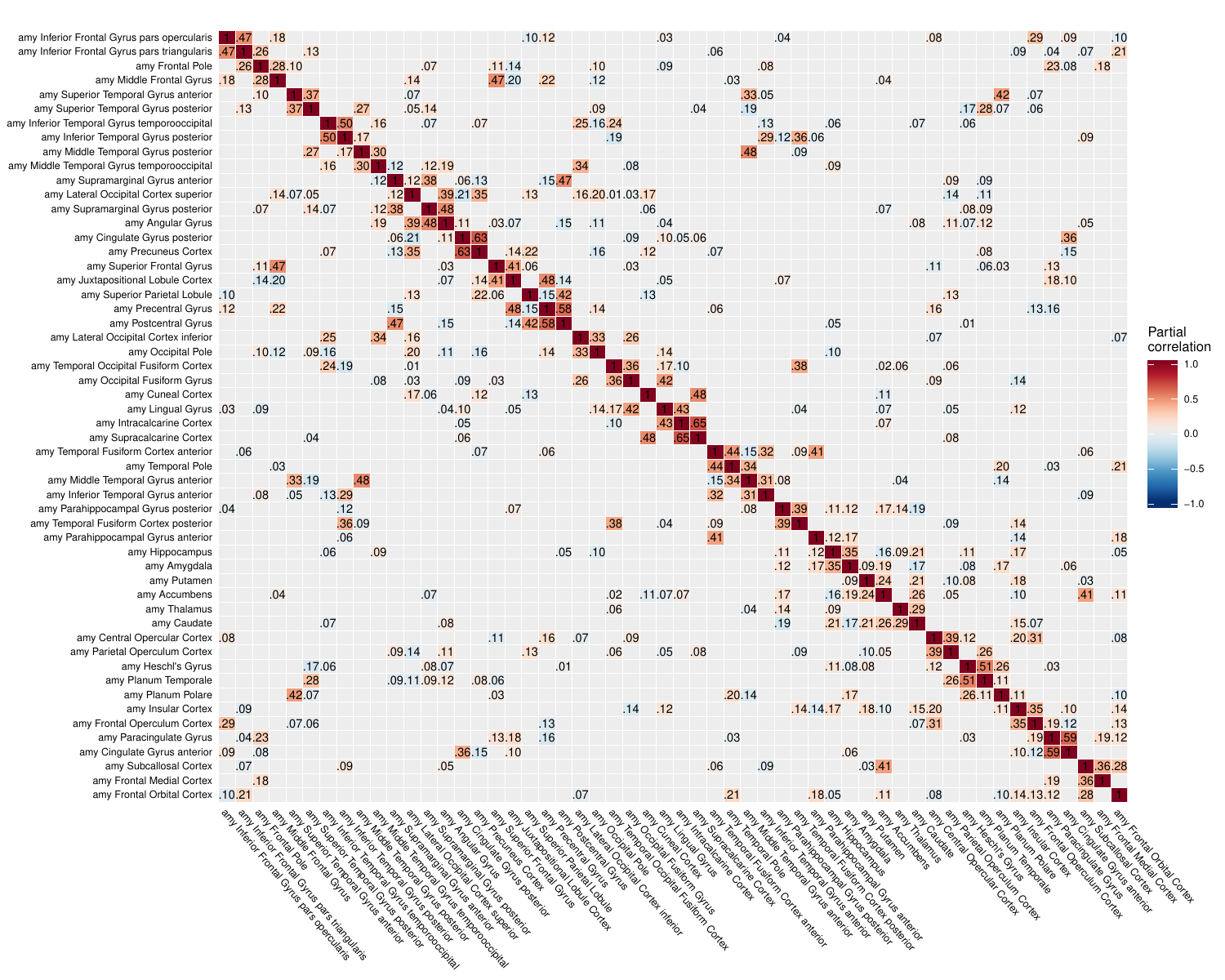}
	\caption{Partial correlation matrix for amyloid-$\beta$ deposition in the left hemisphere estimated for the combined data of EMCI, LMCI and AD patients. Averaged over ten repetitions. Associations of lowest magnitude were not present in all iterations.\\
EMCI/LMCI: early and late amnestic mild cognitive impairment, AD: Alzheimer's dementia, amy: amyloid-$\beta$.}
	\label{fig:parcor-mat-lh-amy-amy}
\end{figure*}

\begin{figure*}[!ht]
	\centering
		\includegraphics[width=\textwidth]{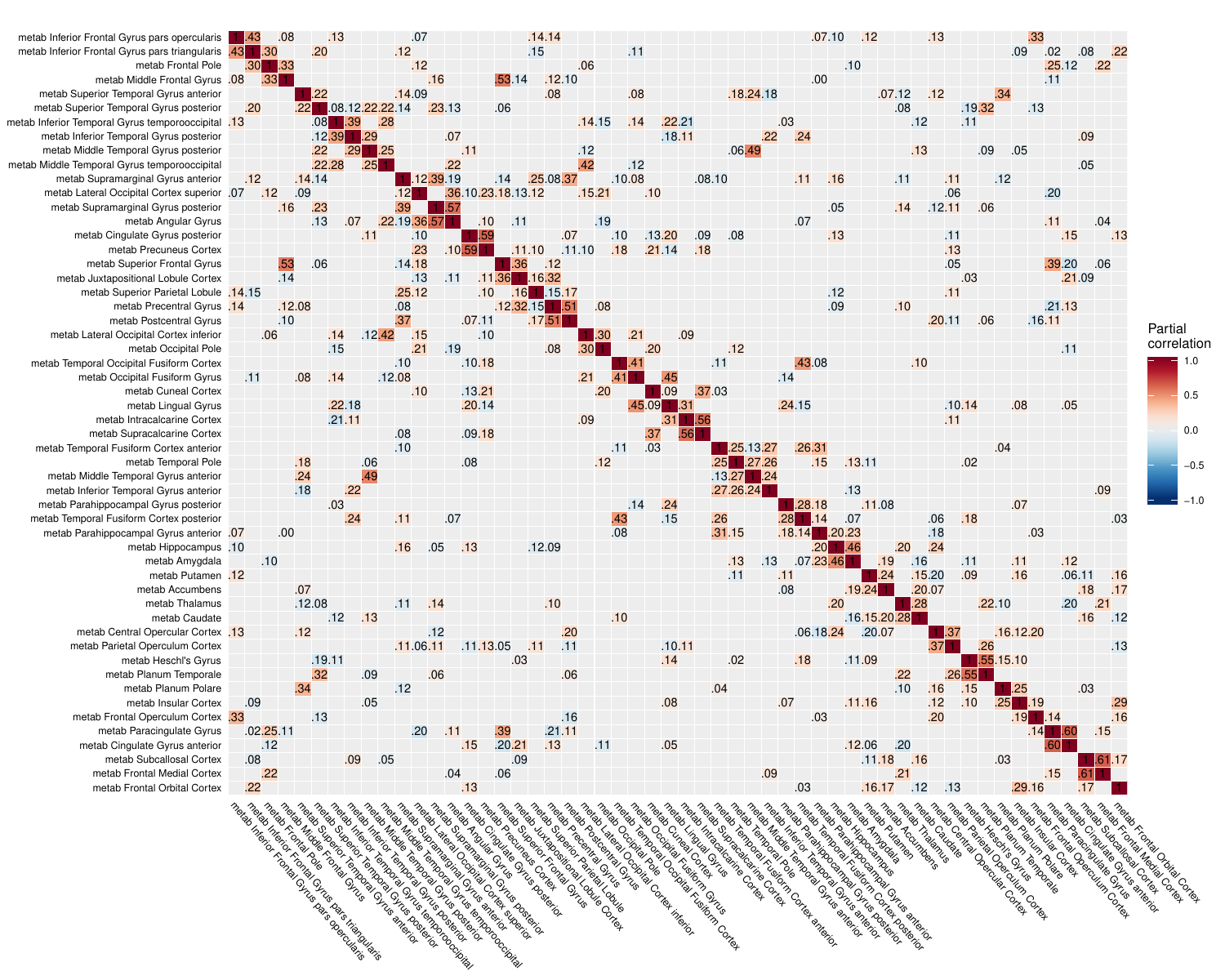}
	\caption{Partial correlation matrix for glucose metabolism in the left hemisphere estimated for the combined data of EMCI, LMCI and AD patients. Averaged over ten repetitions. Associations of lowest magnitude were not present in all iterations.\\
EMCI/LMCI: early and late amnestic mild cognitive impairment, AD: Alzheimer's dementia, metab: glucose metabolism.}
	\label{fig:parcor-mat-lh-fdg-fdg}
\end{figure*}

\begin{figure*}[!ht]
	\centering
		\includegraphics[width=\textwidth]{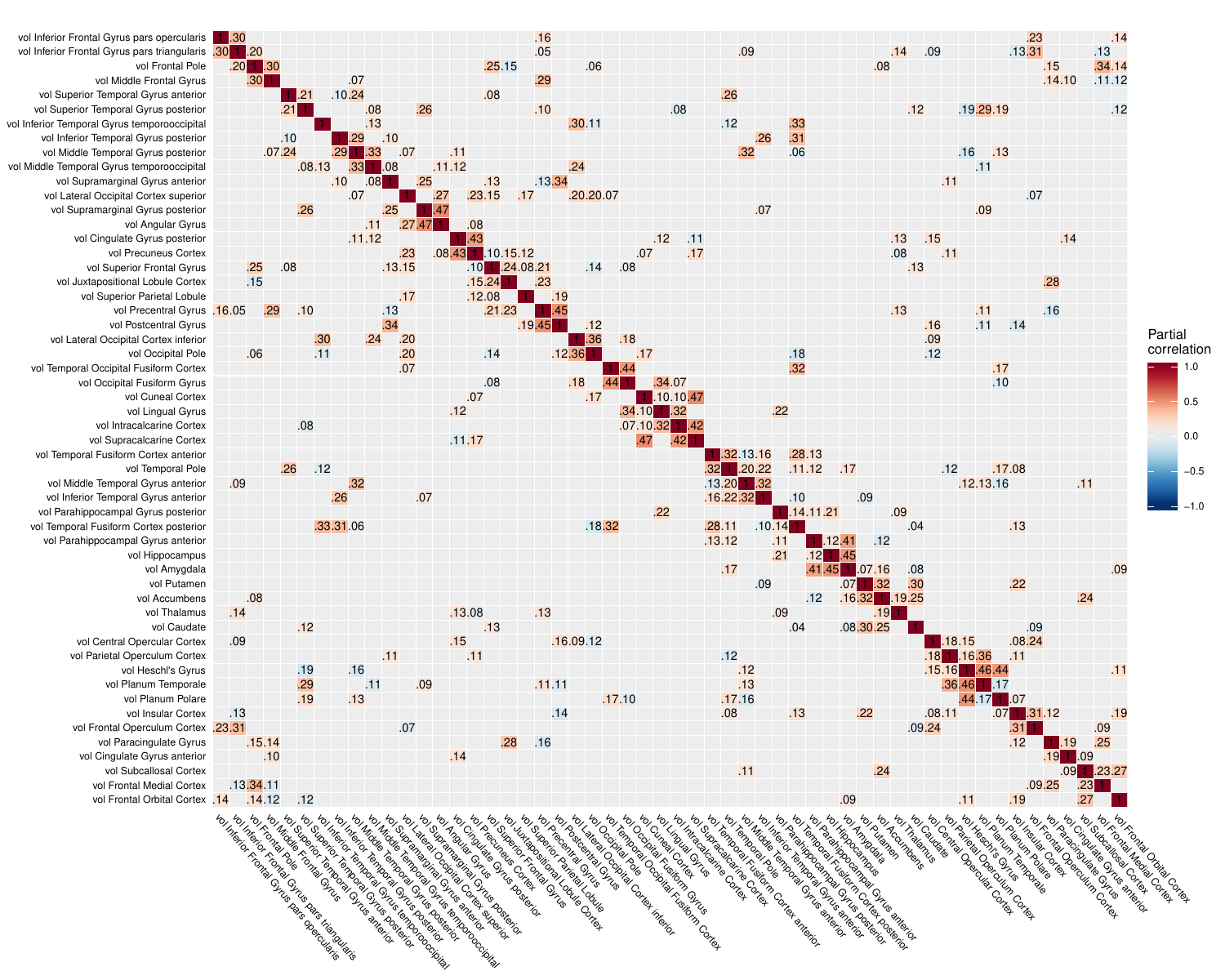}
	\caption{Partial correlation matrix for gray matter volume in the left hemisphere estimated for the combined data of EMCI, LMCI and AD patients. Averaged over ten repetitions. Associations of lowest magnitude were not present in all iterations.\\
EMCI/LMCI: early and late amnestic mild cognitive impairment, AD: Alzheimer's dementia, vol: gray matter volume.}
	\label{fig:parcor-mat-lh-vol-vol}
\end{figure*}

For intra-modal associations, i.e. within the same imaging modality, brain regions directly adjacent to each other formed smaller clusters of high partial correlation around the main diagonal (Figures~\ref{fig:parcor-mat-lh-amy-amy}, \ref{fig:parcor-mat-lh-fdg-fdg}, and \ref{fig:parcor-mat-lh-vol-vol}).
When considering inter-modal associations, i.e. between different imaging modalities, we obtained a consistent pattern of significant positive intra-regional conditional dependency for the pairs amyloid-$\beta$ deposition and metabolism with a mean partial correlation of $\rho=0.21$ for 43 significant associations. 
These are visible as the higher intensities in the diagonal of Supplementary Figure~\ref{sfig:parcor-mat-lh-amy-fdg}.
Between amyloid-$\beta$ and gray matter volume as well as between metabolism and gray matter volume, only few significant intra-regional associations were found (Supplementary Figures \ref{sfig:parcor-mat-lh-fdg-vol} and \ref{sfig:parcor-mat-lh-amy-vol}).

\subsection{Group comparison of the graph structures}

\begin{figure*}[!ht]
	\centering
		\includegraphics[width=0.9\textwidth]{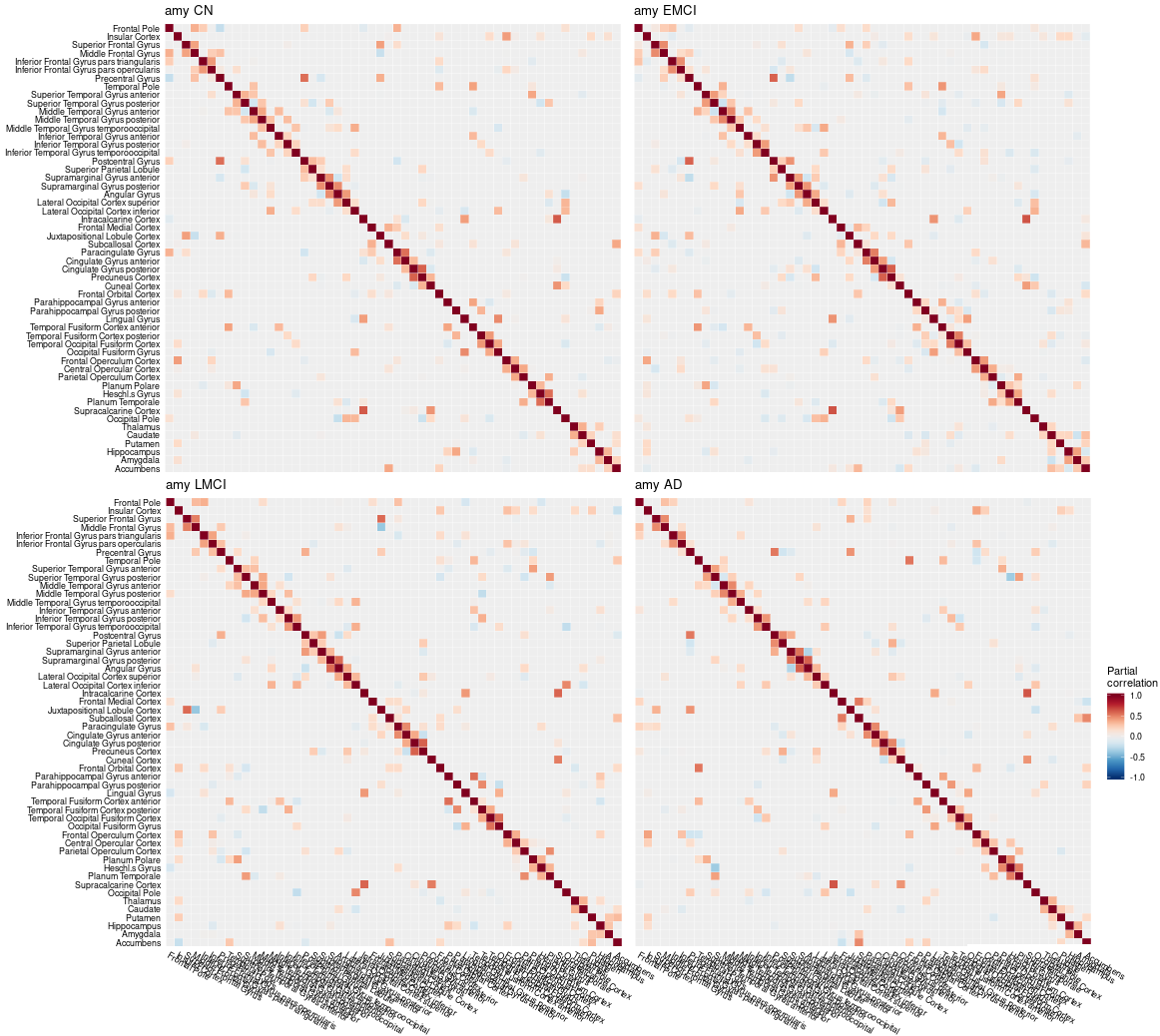}
	\caption{Partial correlation matrix for amyloid-$\beta$ in the left hemisphere by group. Averaged over ten repetitions. Associations of lowest magnitude were not present in all iterations.\\
CN: cognitively healthy elderly controls, EMCI/LMCI: early and late amnestic mild cognitive impairment, AD: Alzheimer's dementia, amy: amyloid-$\beta$.}
	\label{fig:parcor-mat-lh-amy-by-group}
\end{figure*}

\begin{figure*}[!ht]
	\centering
		\includegraphics[width=0.9\textwidth]{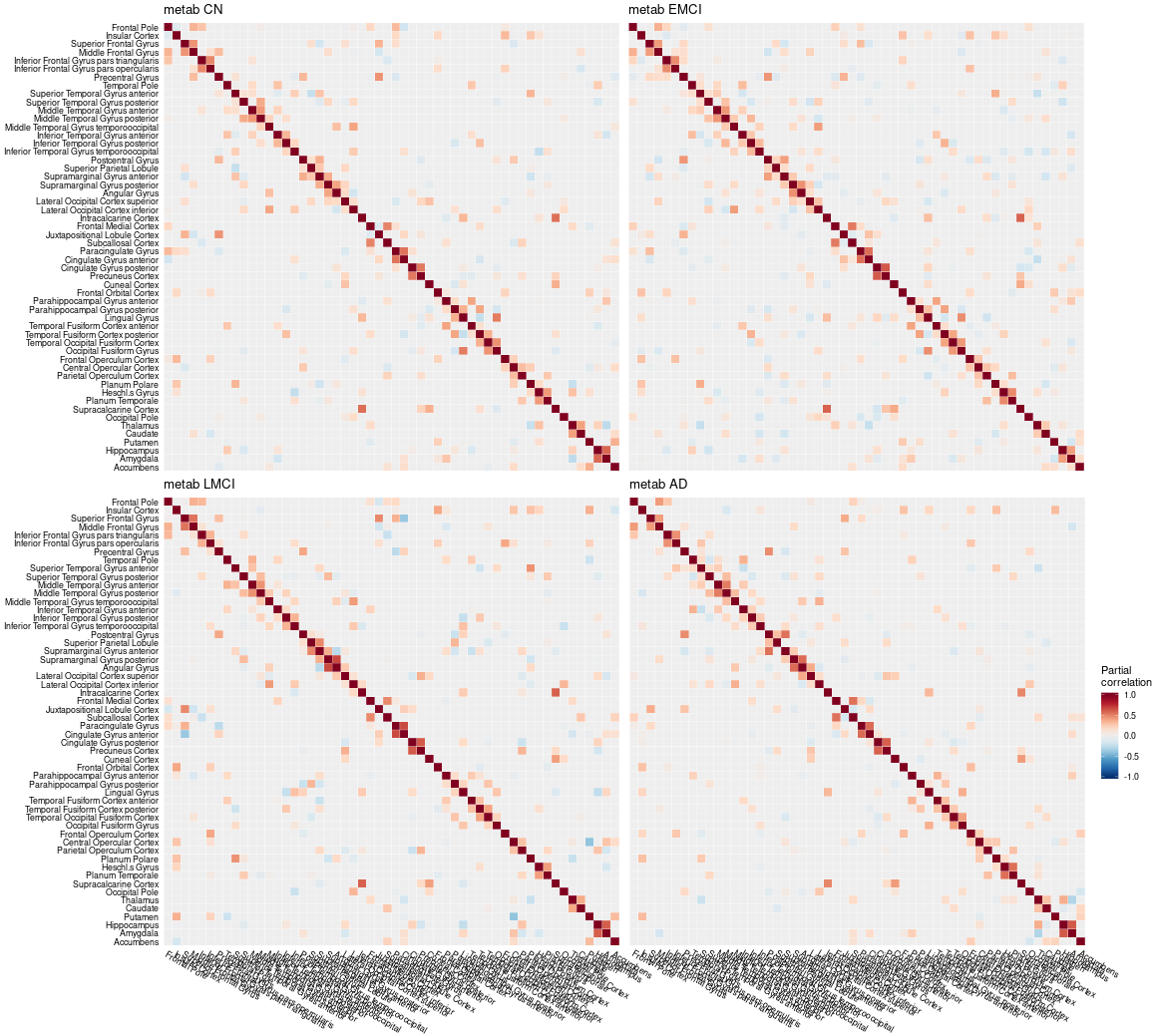}
	\caption{Partial correlation matrix for glucose metabolism in the left hemisphere by group. Averaged over ten repetitions. Associations of lowest magnitude were not present in all iterations.\\
CN: cognitively healthy elderly controls, EMCI/LMCI: early and late amnestic mild cognitive impairment, AD: Alzheimer's dementia, metab: glucose metabolism.}
	\label{fig:parcor-mat-lh-metab-by-group}
\end{figure*}

\begin{figure*}[!ht]
	\centering
		\includegraphics[width=0.9\textwidth]{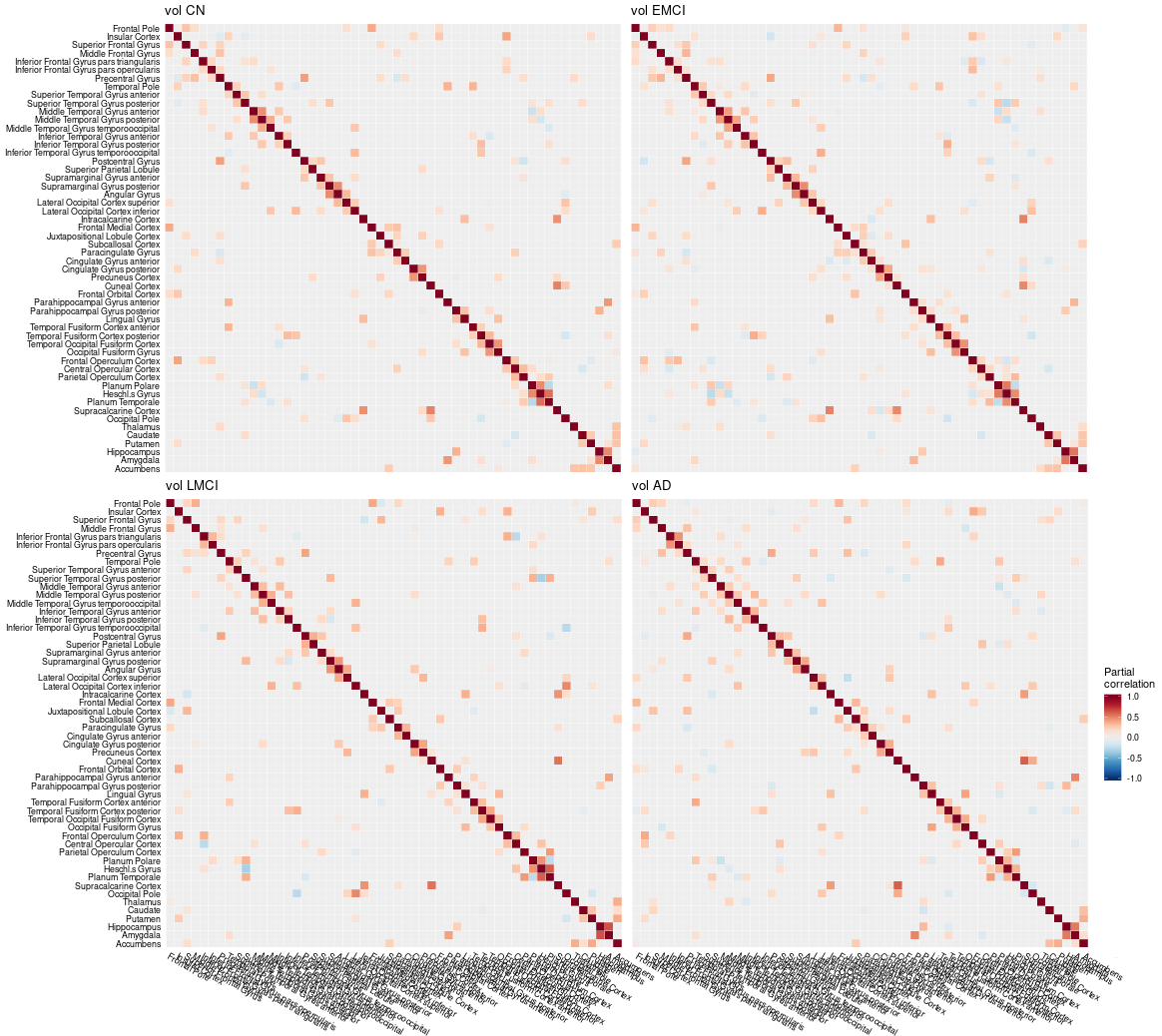}
	\caption{Partial correlation matrix for gray matter volume in the left hemisphere by group. Averaged over ten repetitions. Associations of lowest magnitude were not present in all iterations.\\
CN: cognitively healthy elderly controls, EMCI/LMCI: early and late amnestic mild cognitive impairment, AD: Alzheimer's dementia, vol: gray matter volume.}
	\label{fig:parcor-mat-lh-vol-by-group}
\end{figure*}

\begin{figure*}[!ht]
	\centering
		\includegraphics[width=0.9\textwidth]{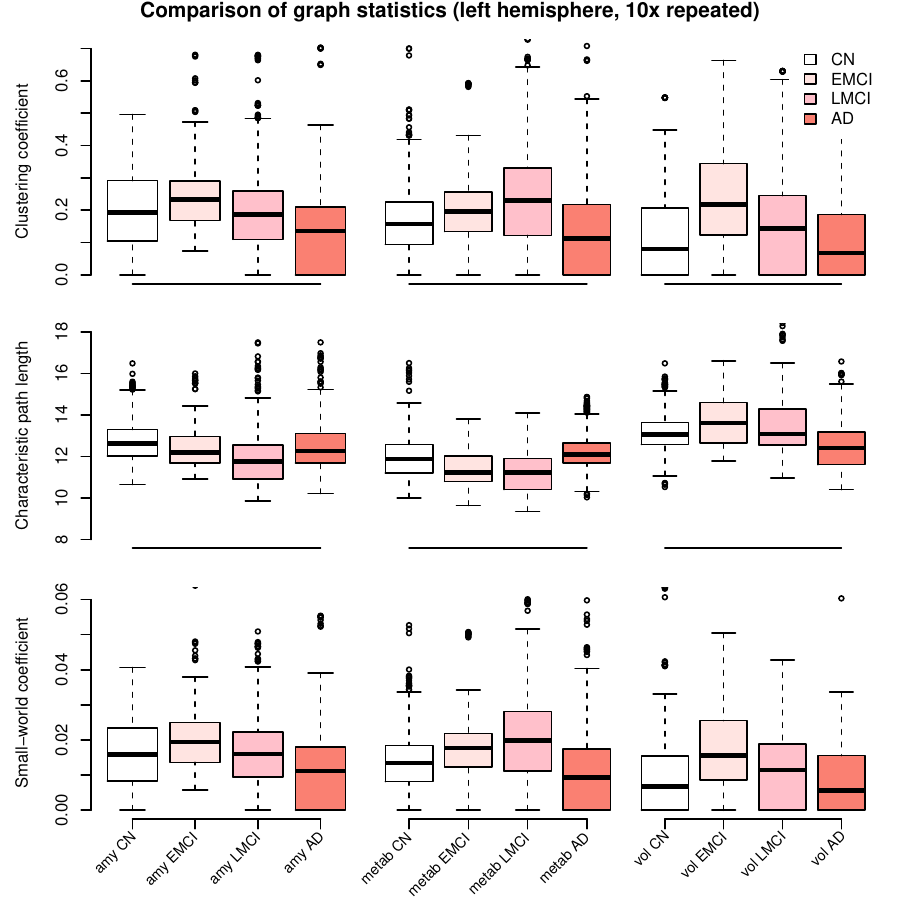}
	\caption{Comparison of graph statistics for the partial correlation matrices of the left hemisphere stratified by diagnostic group and image modality.
	Estimates based on Gaussian graphical models using multimodal neuroimaging data.
	The distribution of the weighted clustering coefficient, characteristic weighted path length, and small-world coefficient for individual brain regions is shown.
	\changed{Boxes display median, first and third quartile of the distributions, and whiskers indicate $\pm1.5 \times$interquartile range.
	All blocks showed significant differences in mean between groups, one-way analysis of variance (ANOVA), $df=215$, $F>4$, $p<0.01$. 
	P-values for Tukey's honest significant difference tests are given in Tab.~\ref{tab:pvals}.}\\
	CN: cognitively healthy elderly controls, EMCI/LMCI: early and late amnestic mild cognitive impairment, AD: Alzheimer's dementia, amy: amyloid-$\beta$, metab: glucose metabolism, vol: gray matter volume.}
	\label{fig:graph_statistics}
\end{figure*}

%
When estimating separate models for each diagnostic group based on the multimodal data, graph structures derived from Pearson and partial correlation matrices (Figures~\ref{fig:parcor-mat-lh-amy-by-group}, \ref{fig:parcor-mat-lh-metab-by-group}, and~\ref{fig:parcor-mat-lh-vol-by-group}) both differed in their density, leading to significant alterations of the clustering coefficient, characteristic path length, and small-world coefficient (Figure~\ref{fig:graph_statistics} and Supplementary Figure~\ref{sfig:graph_stats_Pearson}).
We observed a biphasic trajectory of the graph measures. This means that the clustering coefficient and small world coefficient initially increases when comparing early MCI and CN participants (Figure~\ref{fig:graph_statistics}). When Alzheimer's disease progresses, i.e. in the late MCI and dementia groups, both measures decrease again, with late MCI being approximately on the same level as CN participants (Figure~\ref{fig:graph_statistics}). The characteristic path length showed a similar pattern across groups, but with inverted directionality.
\changed{All blocks showed significant differences in mean between groups, one-way analysis of variance (ANOVA), $df=215$, $F\geq4$, $p<0.01$, $\eta^2 \geq 0.055$. Detailed results are provided in Supplementary Table~\ref{stab:anova}.} P-values for Tukey's honest significant difference tests are provided in Table~\ref{tab:pvals} and Supplementary Table~\ref{stab:pvals2}.
\changed{Graph statistics obtained from the right hemisphere data (Supplementary Figure~\ref{sfig:graph_statistics_righthemi}) were largely consistent with strongest agreement for the characteristic path length metric.}



\section{Discussion}

\subsection{Conditional dependency between brain regions.}
The GGMs estimated the strongest conditional dependencies mainly \emph{within} imaging modalities. We expected adjacent brain regions to form clusters with high inter-cluster similarity for amyloid-$\beta$ deposition (Figure~\ref{fig:parcor-mat-lh-amy-amy}), as it is known to have low variability in spatial distribution and, therefore, is often used as a dichotomic variable after applying a certain threshold to the global amyloid tracer uptake \cite{Chetelat.2013,Landau.2013,Grothe.2017} or as four-stage variable derived from a linear spreading pattern \cite{Grothe.2017,Sakr.2019}.
We also found such clustering patterns for metabolism (Figure~\ref{fig:parcor-mat-lh-fdg-fdg}) and gray matter volume (Figure~\ref{fig:parcor-mat-lh-vol-vol}), matching previous studies on metabolism and gray matter covariance networks based on Pearson correlation \cite{Yao.2010,Carbonell.2016,Pereira.2016} or principal component analysis \cite{Di.2012,Spetsieris.2015,Savio.2017}. Clusters of high covariance have been found in the lateral and medial parietal lobe, lateral frontal lobe, and lateral and medial temporal lobe, and had been associated with simultaneous growth during brain development, functional co-activation, and axonal connectivity in the literature \cite{Gong.2012,AlexanderBloch.2013}.

Our analyses yielded only few and relatively weak associations \emph{between} different modalities (Figure~\ref{sfig:parcor-mat-lh-amy-fdg}-\ref{sfig:parcor-mat-lh-amy-vol}), except for the direct intra-regional dependency between amyloid-$\beta$ and metabolism as well as between amyloid and gray matter volume (diagonal of Figure~\ref{sfig:parcor-mat-lh-amy-fdg} and Figure~\ref{sfig:parcor-mat-lh-amy-vol}), which matched our previous analysis with six selected regions of interest \cite{Dyrba.2017}. 
The positive dependency between amyloid-$\beta$ and metabolism was strongest in the early MCI group and matches previous results for partial correlation obtained from linear regression models \cite{Altmann.2015}. This previous study reported a markedly reduced number and strength of negative associations between regional amyloid-$\beta$ and metabolism when correcting for global amyloid load. They concluded that the negative association between amyloid deposition and metabolism is more related to the global amyloid level than to the distinct regional level.
The pattern of intra-regional dependency between amyloid-$\beta$ and metabolism as well as between amyloid-$\beta$ and gray matter volume was strongest in the early MCI group, which could refer to the early phase of the disease and, therefore, a high variation in regional amyloid-$\beta$ deposition and a strong contribution of the amyloid level on both metabolism and volume \cite{Drzezga.2011,Carbonell.2016}.
Notably, conditional dependencies between metabolism and volume were obtained only for few regions including hippocampus and putamen, but not for other expected regions such as posterior cingulate cortex \cite{Teipel.2016} (Supplementary Figure~\ref{sfig:parcor-mat-lh-fdg-vol}).

\subsection{Alterations of graph measures}

Various studies reported a network disruption of AD in comparison to cognitively healthy controls for gray matter volume \cite{He.2008,Yao.2010,Li.2012,Tijms.2013,John.2017} and glucose metabolism \cite{Morbelli.2012,Titov.2017}, and intermediate levels for volume in MCI \cite{Yao.2010,Pereira.2016}; which we could replicate in our sample (Supplementary Figure \ref{sfig:graph_stats_Pearson}).
However, it has to be noted that for Pearson correlation matrices usually high thresholds are applied to obtain sparser graphs.
Chung et al. \cite{Chung.2016} and Voevodskaya et al. \cite{Voevodskaya.2017}  reported a high influence of the selected graph density threshold on the graph measures, leading to divergent increases and decreases of the global clustering coefficient metric.
To circumvent such problems, we used weighted versions of the graph measures \cite{Rubinov.2010} and proposed GGMs to obtain sparse conditional dependency matrices. Our results suggest that graph statistics for regional dependency networks follow a biphasic trajectory in the course of AD, a pattern that was recently also reported for cortical thinning and mean diffusivity \cite{Montal.2017} and resting-state fMRI connectivity \cite{Schultz.2017}.

In the current study, the EMCI group displayed the strongest alterations of network structure with an increase of the clustering coefficient, which may relate to the process of amyloid accumulation taking place in several regions simultaneously in this group increasing the intra-cluster correlation. 
For amyloid-$\beta$ and volume, LMCI subjects showed a clustering coefficient and small-world coefficient comparable to controls, in contrast to metabolism, where this group showed strongest deviation from the other groups (Table~\ref{tab:pvals}).
The lowest alterations of graph measures were obtained for the gray matter network. 

 
GGMs were recently applied as clustering algorithm for brain networks in a few other single-modality applications. De Vos et al. \cite{Vos.2017} found them useful for increasing group separation between AD and controls compared to classical Pearson correlation networks in resting-state functional connectivity.
Titov et al. \cite{Titov.2017} compared metabolic networks for the differential diagnosis between AD and frontotemporal lobar degeneration (FTLD). They also proposed an algorithm to estimate if an individual subject shows a more AD or FTLD pattern of regional metabolism.
Munilla et al. \cite{Munilla.2017} systematically evaluated the influence of the number of subjects and the regularization strength on the GGM stability and graph structure. They found that the estimated GGM graph structure and small-world coefficient converged to a stable level when including 40 or more subjects in their study sample. For regularization-based approximation of GGMs, they showed that the probability of an edge to exist in the estimated graph structure almost linearly corresponds to the magnitude of their partial correlation. Thus, this finding confirms our initial decision, that sampling-based Bayesian estimation of the graph structure might be more useful for detecting even low associations.

\subsection{Limitations}

It has to be noted that our methodological framework can currently only be applied as a group statistic but not for individual subjects. Therefore, GGMs can be used for exploratory analyses as alternative to Pearson correlation networks, and may aid generating new hypotheses about the interrelation of clinical variables or feature selection. Then, derived hypotheses can be validated using classical statistical methods such as regression or mediation analysis.

\changed{Another limitation is the high uncertainty in the statistical model to estimate the partial correlations. This is due to the theoretically hard problem of matrix inversion on the one hand, and due to the high number of possible graph edges in comparison to the sample size on the other hand.
Thus, the model might be fragile with respect to the obtained values and requires large training samples to get stable results. 
Here, we repeated the model estimation on the whole data for ten times to observe the effect on model stability, which was yielding largely consistent results for strong links with high partial correlation, but getting more variable for weaker links with low partial correlation.
Replicating the results using the right hemisphere data also yielded largely consistent results with highest agreement for the characteristic path length metric. Apparent deviation in clustering coefficient and consequently in small-world coefficient (=ratio of both) might be explained by the asymmetry of the brain and the lateralization reported for Alzheimer’s disease in the literature (e.g. stronger left hippocampus atrophy in ADNI) \cite{Grothe.2016b,Weise.2018}.
However, our findings still need to be replicated in independent cohorts.}

We observed a saturation of the conditional dependency network when adding many variables. This means, the model parameters might strongly change when having only few variables in the model and adding another variable; in contrast to very stable estimates of larger models with dozens of variables, which are hardly altered when adding another variable.
Actually, this problem is well-known for linear regression models and related to multicollinearity in the data \cite{Obrien.2007,Dormann.2013,Teipel.2015b}.
Recent developments in stochastic block models may help to overcome these limitations, as they try to infer the underlying clustering block structure and separately estimate statistical associations within and between clusters \cite{Sun.2014,Hosseini.2016}.

\subsection{Conclusion}

We applied GGMs to assess inter-modal and inter-regional dependencies of high-dimensional multimodal neuroimaging data of AD-related brain alterations.
Our results showed that conditional dependency networks estimated by GGMs provide useful information within imaging modalities and could be used as alternative to Pearson-correlation networks.
Nonetheless, GGMs did not detect some expected associations between modalities and, therefore, may have limited applicability for large-scale data with dozens of variables.

\section*{Conflict of interest}
The authors declare no conflict of interest with respect to this study.

\section*{Author contributions}
MD, RM, TM and ST designed the study. MG and MD preprocessed the imaging data. MD and RM conducted the statistical analyses. MG, TK, ST aided in interpreting the results. MD drafted the first version of the manuscript. All authors revised the manuscript and contributed to the final version.

\section*{Funding}
This project was supported by the Rostock Massive Data Research Facility (RMDRF) funded by the German Research Foundation (DFG), grant number FKZ INST 264/128-1 FUGG.

\section*{Acknowledgments}
Data collection and sharing for this project was funded by the Alzheimer's Disease Neuroimaging Initiative (ADNI) (National Institutes of Health Grant U01 AG024904). The ADNI was launched in 2003 by the National Institute on Aging (NIA), the National Institute of Biomedical Imaging and Bioengineering (NIBIB), the Food and Drug Administration (FDA), private pharmaceutical companies and non-profit organizations, as a \$60 million, 5-year public-private partnership. ADNI is funded by the National Institute on Aging, the National Institute of Biomedical Imaging and Bioengineering, and through generous contributions from the following: Alzheimer's Association; Alzheimer's Drug Discovery Foundation; Araclon Biotech; BioClinica, Inc.; Biogen Idec Inc.; Bristol-Myers Squibb Company; CereSpir, Inc.; Eisai Inc.; Elan Pharmaceuticals, Inc.; Eli Lilly and Company; EuroImmun; F. Hoffmann-La Roche Ltd and its affiliated company Genentech, Inc.; Fujirebio; GE Healthcare; IXICO Ltd.; Janssen Alzheimer Immunotherapy Research \& Development, LLC.; Johnson \& Johnson Pharmaceutical Research \& Development LLC.; Lumosity; Lundbeck; Merck \& Co., Inc.; Meso Scale Diagnostics, LLC.; NeuroRx Research; Neurotrack Technologies; Novartis Pharmaceuticals Corporation; Pfizer Inc.; Piramal Imaging; Servier; Takeda Pharmaceutical Company; and Transition Therapeutics. The Canadian Institute of Health Research is providing funds to support ADNI clinical sites in Canada. Private sector contributions are facilitated by the Foundation for the National Institutes of Health (www.fnih.org). The grantee organization is the Northern California Institute for Research and Education, and the study is coordinated by the Alzheimer's Disease Cooperative Study at the University of California, San Diego. ADNI data are disseminated by the Laboratory for Neuroimaging at the University of Southern California.

\section*{Data availability statement}
MRI and PET data being used in this study can be retrieved from ADNI (\url{http://adni.loni.usc.edu/data-samples/access-data/}). Processed imaging data and extracted regional mean values are available from the corresponding authors upon request. The R package BDgraph can be downloaded from CRAN (\url{https://cran.r-project.org/web/packages/BDgraph}) or GitHub (\url{https://github.com/cran/BDgraph}).


\bibliography{AlzheimerBrain}

\begin{thebibliography}{}

\bibitem[Alexander-Bloch et~al., 2013]{AlexanderBloch.2013}
Alexander-Bloch, A., Raznahan, A., Bullmore, E., and Giedd, J. (2013).
\newblock The convergence of maturational change and structural covariance in
  human cortical networks.
\newblock {\em The Journal of neuroscience : the official journal of the
  Society for Neuroscience}, 33(7):2889--2899.

\bibitem[Altmann et~al., 2015]{Altmann.2015}
Altmann, A., Ng, B., Landau, S.~M., Jagust, W.~J., and Greicius, M.~D. (2015).
\newblock Regional brain hypometabolism is unrelated to regional amyloid plaque
  burden.
\newblock {\em Brain}, 138(Pt 12):3734--3746.

\bibitem[Bontempi and Flauder, 2015]{Bontempi.2015}
Bontempi, G. and Flauder, M. (2015).
\newblock From dependency to causality: A machine learning approach.
\newblock {\em Journal of Machine Learning Research}, 16(74):2437--2457.

\bibitem[Buckner et~al., 2009]{Buckner.2009}
Buckner, R.~L., Sepulcre, J., Talukdar, T., Krienen, F.~M., Liu, H., Hedden,
  T., Andrews-Hanna, J.~R., Sperling, R.~A., and Johnson, K.~A. (2009).
\newblock Cortical hubs revealed by intrinsic functional connectivity: mapping,
  assessment of stability, and relation to alzheimer's disease.
\newblock {\em The Journal of Neuroscience}, 29(6):1860--1873.

\bibitem[Buckner et~al., 2005]{Buckner.2005}
Buckner, R.~L., Snyder, A.~Z., Shannon, B.~J., LaRossa, G., Sachs, R., Fotenos,
  A.~F., Sheline, Y.~I., Klunk, W.~E., Mathis, C.~A., Morris, J.~C., and
  Mintun, M.~A. (2005).
\newblock Molecular, structural, and functional characterization of
  {A}lzheimer's disease: Evidence for a relationship between default activity,
  amyloid, and memory.
\newblock {\em The Journal of Neuroscience}, 25(34):7709--7717.

\bibitem[Cai et~al., 2013]{Cai.2013}
Cai, T.~T., Li, H., Liu, W., and Xie, J. (2013).
\newblock Covariate-adjusted precision matrix estimation with an application in
  genetical genomics.
\newblock {\em Biometrika}, 100(1):139--156.

\bibitem[Carbonell et~al., 2016]{Carbonell.2016}
Carbonell, F., Zijdenbos, A.~P., McLaren, D.~G., Iturria-Medina, Y., and
  Bedell, B.~J. (2016).
\newblock Modulation of glucose metabolism and metabolic connectivity by
  beta-amyloid.
\newblock {\em Journal of Cerebral Blood Flow {\&} Metabolism},
  36(12):2058--2071.

\bibitem[Chang et~al., 2015]{Chang.2015}
Chang, Y.-T., Huang, C.-W., Chang, Y.-H., Chen, N.-C., Lin, K.-J., Yan, T.-C.,
  Chang, W.-N., Chen, S.-F., Lui, C.-C., Lin, P.-H., and Chang, C.-C. (2015).
\newblock Amyloid burden in the hippocampus and default mode network:
  Relationships with gray matter volume and cognitive performance in mild stage
  {A}lzheimer disease.
\newblock {\em Medicine}, 94(16):e763.

\bibitem[Ch{\'e}telat et~al., 2013]{Chetelat.2013}
Ch{\'e}telat, G., {La Joie}, R., Villain, N., Perrotin, A., de~{La Sayette},
  V., Eustache, F., and Vandenberghe, R. (2013).
\newblock Amyloid imaging in cognitively normal individuals, at-risk
  populations and preclinical alzheimer's disease.
\newblock {\em NeuroImage. Clinical}, 2:356--365.

\bibitem[Chung et~al., 2016]{Chung.2016}
Chung, J., Yoo, K., Kim, E., Na, D.~L., and Jeong, Y. (2016).
\newblock Glucose metabolic brain networks in early-onset vs. late-onset
  alzheimer's disease.
\newblock {\em Frontiers in aging neuroscience}, 8:159.

\bibitem[de~Vos et~al., 2017]{Vos.2017}
de~Vos, F., Koini, M., Schouten, T.~M., Seiler, S., {van der Grond}, J.,
  Lechner, A., Schmidt, R., de~Rooij, M., and Rombouts, S. A. R.~B. (2017).
\newblock A comprehensive analysis of resting state fmri measures to classify
  individual patients with alzheimer's disease.
\newblock {\em NeuroImage}, 167:62--72.

\bibitem[Desikan et~al., 2006]{Desikan.2006}
Desikan, R.~S., S{\'e}gonne, F., Fischl, B., Quinn, B.~T., Dickerson, B.~C.,
  Blacker, D., Buckner, R.~L., Dale, A.~M., Maguire, R.~P., Hyman, B.~T.,
  Albert, M.~S., and Killiany, R.~J. (2006).
\newblock An automated labeling system for subdividing the human cerebral
  cortex on {MRI} scans into gyral based regions of interest.
\newblock {\em NeuroImage}, 31(3):968--980.

\bibitem[Di and Biswal, 2012]{Di.2012}
Di, X. and Biswal, B.~B. (2012).
\newblock Metabolic brain covariant networks as revealed by fdg-pet with
  reference to resting-state fmri networks.
\newblock {\em Brain connectivity}, 2(5):275--283.

\bibitem[Dormann et~al., 2013]{Dormann.2013}
Dormann, C.~F., Elith, J., Bacher, S., Buchmann, C., Carl, G., Carr{\'e}, G.,
  Marqu{\'e}z, J. R.~G., Gruber, B., Lafourcade, B., Leit{\~a}o, P.~J.,
  M{\"u}nkem{\"u}ller, T., McClean, C., Osborne, P.~E., Reineking, B.,
  Schr{\"o}der, B., Skidmore, A.~K., Zurell, D., and Lautenbach, S. (2013).
\newblock Collinearity: A review of methods to deal with it and a simulation
  study evaluating their performance.
\newblock {\em Ecography}, 36(1):27--46.

\bibitem[Drzezga et~al., 2011]{Drzezga.2011}
Drzezga, A., Becker, J.~A., {van Dijk}, K. R.~A., Sreenivasan, A., Talukdar,
  T., Sullivan, C., Schultz, A.~P., Sepulcre, J., Putcha, D., Greve, D.,
  Johnson, K.~A., and Sperling, R.~A. (2011).
\newblock Neuronal dysfunction and disconnection of cortical hubs in
  non-demented subjects with elevated amyloid burden.
\newblock {\em Brain}, 134(Pt 6):1635--1646.

\bibitem[Dyrba et~al., 2017]{Dyrba.2017}
Dyrba, M., Grothe, M.~J., Mohammadi, A., Binder, H., Kirste, T., and Teipel,
  S.~J. (2017).
\newblock Comparison of different hypotheses regarding the spread of
  {A}lzheimer's disease using {M}arkov random fields and multimodal imaging.
\newblock {\em Journal of Alzheimer's disease}.

\bibitem[Gong et~al., 2012]{Gong.2012}
Gong, G., He, Y., Chen, Z.~J., and Evans, A.~C. (2012).
\newblock Convergence and divergence of thickness correlations with diffusion
  connections across the human cerebral cortex.
\newblock {\em NeuroImage}, 59(2):1239--1248.

\bibitem[Gonzalez-Escamilla et~al., 2017]{GonzalezEscamilla.2017}
Gonzalez-Escamilla, G., Lange, C., Teipel, S., Buchert, R., and Grothe, M.~J.
  (2017).
\newblock {PETPVE12: An SPM toolbox for Partial Volume Effects correction in
  brain PET -- Application to amyloid imaging with AV45-PET}.
\newblock {\em {NeuroImage}}, 147:669--677.

\bibitem[Grothe et~al., 2013]{Grothe.2013}
Grothe, M., Heinsen, H., and Teipel, S. (2013).
\newblock Longitudinal measures of cholinergic forebrain atrophy in the
  transition from healthy aging to alzheimer's disease.
\newblock {\em Neurobiology of Aging}, 34(4):1210--1220.

\bibitem[Grothe et~al., 2017]{Grothe.2017}
Grothe, M.~J., Barthel, H., Sepulcre, J., Dyrba, M., Sabri, O., and Teipel,
  S.~J. (2017).
\newblock In vivo staging of regional amyloid deposition.
\newblock {\em Neurology}.

\bibitem[Grothe et~al., 2016]{Grothe.2016a}
Grothe, M.~J., Heinsen, H., Amaro, E., Grinberg, L.~T., and Teipel, S.~J.
  (2016).
\newblock Cognitive correlates of basal forebrain atrophy and associated
  cortical hypometabolism in mild cognitive impairment.
\newblock {\em Cerebral Cortex}, 26(6):2411--2426.

\bibitem[Grothe and Teipel, 2016]{Grothe.2016b}
Grothe, M.~J. and Teipel, S.~J. (2016).
\newblock Spatial patterns of atrophy, hypometabolism, and amyloid deposition
  in {A}lzheimer's disease correspond to dissociable functional brain networks.
\newblock {\em Human Brain Mapping}, 37(1):35--53.

\bibitem[Hastie et~al., 2013]{Hastie.2013}
Hastie, T.~J., Tibshirani, R.~J., and Friedman, J.~H. (2013).
\newblock {\em The elements of statistical learning: Data mining, inference,
  and prediction}.
\newblock Springer series in statistics. Springer, New York, NY, 2. ed.
  edition.

\bibitem[He et~al., 2008]{He.2008}
He, Y., Chen, Z., and Evans, A. (2008).
\newblock Structural insights into aberrant topological patterns of large-scale
  cortical networks in alzheimer's disease.
\newblock {\em The Journal of neuroscience : the official journal of the
  Society for Neuroscience}, 28(18):4756--4766.

\bibitem[Hlinka et~al., 2017]{Hlinka.2017}
Hlinka, J., Hartman, D., Jajcay, N., Tomeček, D., Tintěra, J., and Paluš, M.
  (2017).
\newblock Small-world bias of correlation networks: From brain to climate.
\newblock {\em Chaos}, 27(3):035812.

\bibitem[Hosseini and Lee, 2016]{Hosseini.2016}
Hosseini, M.~J. and Lee, S.-I. (2016).
\newblock Learning sparse gaussian graphical models with overlapping blocks.
\newblock In Lee, D.~D., Sugiyama, M., Luxburg, U.~V., Guyon, I., and Garnett,
  R., editors, {\em Advances in Neural Information Processing Systems 29},
  pages 3808--3816. Curran Associates, Inc.

\bibitem[Iturria-Medina et~al., 2017]{IturriaMedina.2017}
Iturria-Medina, Y., Carbonell, F.~M., Sotero, R.~C., Chouinard-Decorte, F., and
  Evans, A.~C. (2017).
\newblock Multifactorial causal model of brain (dis)organization and
  therapeutic intervention: Application to {A}lzheimer's disease.
\newblock {\em NeuroImage}, 152:60--77.

\bibitem[John et~al., 2017]{John.2017}
John, M., Ikuta, T., and Ferbinteanu, J. (2017).
\newblock Graph analysis of structural brain networks in alzheimer's disease:
  Beyond small world properties.
\newblock {\em Brain structure {\&} function}, 222(2):923--942.

\bibitem[Kljajevic et~al., 2014]{Kljajevic.2014}
Kljajevic, V., Grothe, M.~J., Ewers, M., and Teipel, S. (2014).
\newblock Distinct pattern of hypometabolism and atrophy in preclinical and
  predementia {A}lzheimer's disease.
\newblock {\em Neurobiology of Aging}, 35(9):1973--1981.

\bibitem[Koller and Friedman, 2009]{Koller.2009}
Koller, D. and Friedman, N. (2009).
\newblock {\em Probabilistic graphical models: Principles and techniques}.
\newblock Adaptive computation and machine learning. {MIT Press}, Cambridge,
  MA.

\bibitem[{La Joie} et~al., 2012]{LaJoie.2012}
{La Joie}, R., Perrotin, A., Barr{\'e}, L., Hommet, C., M{\'e}zenge, F.,
  Ibazizene, M., Camus, V., Abbas, A., Landeau, B., Guilloteau, D., de~{La
  Sayette}, V., Eustache, F., Desgranges, B., and Ch{\'e}telat, G. (2012).
\newblock Region-specific hierarchy between atrophy, hypometabolism, and
  $\beta$-amyloid (a$\beta$) load in {A}lzheimer's disease dementia.
\newblock {\em The Journal of Neuroscience}, 32(46):16265--16273.

\bibitem[Landau et~al., 2013]{Landau.2013}
Landau, S.~M., Breault, C., Joshi, A.~D., Pontecorvo, M., Mathis, C.~A.,
  Jagust, W.~J., and Mintun, M.~A. (2013).
\newblock Amyloid-beta imaging with pittsburgh compound b and florbetapir:
  Comparing radiotracers and quantification methods.
\newblock {\em Journal of Nuclear Medicine}, 54(1):70--77.

\bibitem[Lauritzen, 1996]{Lauritzen.1996}
Lauritzen, S.~L. (1996).
\newblock {\em Graphical models}, volume~17 of {\em Oxford statistical science
  series}.
\newblock {Clarendon Press}, Oxford.

\bibitem[Li et~al., 2012]{Li.2012}
Li, Y., Wang, Y., Wu, G., Shi, F., Zhou, L., Lin, W., and Shen, D. (2012).
\newblock Discriminant analysis of longitudinal cortical thickness changes in
  {A}lzheimer's disease using dynamic and network features.
\newblock {\em Neurobiology of aging}, 33(2):427.e15--30.

\bibitem[Madigan et~al., 1996]{Madigan.1996}
Madigan, D., Raftery, A.~E., Volinsky, C., and Hoeting, J. (1996).
\newblock Bayesian model averaging.
\newblock In {\em Proceedings of the AAAI Workshop on Integrating Multiple
  Learned Models, Portland, OR}, pages 77--83.

\bibitem[Meinshausen and Bühlmann, 2006]{Meinshausen.2006}
Meinshausen, N. and Bühlmann, P. (2006).
\newblock High-dimensional graphs and variable selection with the lasso.
\newblock {\em Annals of Statistics}, 34(3):1436--1462.

\bibitem[Mohammadi and Wit, 2015]{Mohammadi.2015}
Mohammadi, A. and Wit, E.~C. (2015).
\newblock {B}ayesian structure learning in sparse {G}aussian graphical models.
\newblock {\em Bayesian Analysis}, 10(1):109--138.

\bibitem[Mohammadi and Wit, 2019]{Mohammadi.2019}
Mohammadi, R. and Wit, E.~C. (2019).
\newblock {BDgraph} : An {R} package for {B}ayesian structure learning in
  graphical models.
\newblock {\em Journal of Statistical Software}, 89(3).

\bibitem[Montal et~al., 2017]{Montal.2017}
Montal, V., Vilaplana, E., Alcolea, D., Pegueroles, J., Pasternak, O.,
  González-Ortiz, S., Clarimón, J., Carmona-Iragui, M., Illán-Gala, I.,
  Morenas-Rodríguez, E., Ribosa-Nogué, R., Sala, I., Sánchez-Saudinós,
  M.-B., García-Sebastian, M., Villanúa, J., Izagirre, A., Estanga, A.,
  Ecay-Torres, M., Iriondo, A., Clerigue, M., Tainta, M., Pozueta, A.,
  González, A., Martínez-Heras, E., Llufriu, S., Blesa, R., Sanchez-Juan, P.,
  Martínez-Lage, P., Lleó, A., and Fortea, J. (2017).
\newblock Cortical microstructural changes along the {A}lzheimer's disease
  continuum.
\newblock {\em Alzheimer's {\&} Dementia}.

\bibitem[Morbelli et~al., 2012]{Morbelli.2012}
Morbelli, S., Drzezga, A., Perneczky, R., Frisoni, G.~B., Caroli, A., {van
  Berckel}, B. N.~M., Ossenkoppele, R., Guedj, E., Didic, M., Brugnolo, A.,
  Sambuceti, G., Pagani, M., Salmon, E., and Nobili, F. (2012).
\newblock Resting metabolic connectivity in prodromal alzheimer's disease. a
  european alzheimer disease consortium (eadc) project.
\newblock {\em Neurobiology of Aging}, 33(11):2533--2550.

\bibitem[Munilla et~al., 2017]{Munilla.2017}
Munilla, J., Ortiz, A., Górriz, J.~M., and Ramírez, J. (2017).
\newblock Construction and analysis of weighted brain networks from sice for
  the study of alzheimer's disease.
\newblock {\em Frontiers in Neuroinformatics}, 11:19.

\bibitem[Mårtensson et~al., 2018]{Martensson.2018}
Mårtensson, G., Pereira, J.~B., Mecocci, P., Vellas, B., Tsolaki, M.,
  Kłoszewska, I., Soininen, H., Lovestone, S., Simmons, A., Volpe, G., and
  Westman, E. (2018).
\newblock Stability of graph theoretical measures in structural brain networks
  in {A}lzheimer’s disease.
\newblock {\em Scientific Reports}, 8(1):11592.

\bibitem[Müller-Gärtner et~al., 1992]{MullerGartner.1992}
Müller-Gärtner, H.~W., Links, J.~M., Prince, J.~L., Bryan, R.~N., McVeigh,
  E., Leal, J.~P., Davatzikos, C., and Frost, J.~J. (1992).
\newblock {Measurement of Radiotracer Concentration in Brain Gray Matter Using
  Positron Emission Tomography: MRI-Based Correction for Partial Volume
  Effects}.
\newblock {\em {Journal of Cerebral Blood Flow {\&} Metabolism}},
  12(4):571--583.

\bibitem[O'brien, 2007]{Obrien.2007}
O'brien, R.~M. (2007).
\newblock A caution regarding rules of thumb for variance inflation factors.
\newblock {\em Quality {\&} Quantity}, 41(5):673--690.

\bibitem[Onnela et~al., 2005]{Onnela.2005}
Onnela, J.-P., Saram\"aki, J., Kert\'esz, J., and Kaski, K. (2005).
\newblock Intensity and coherence of motifs in weighted complex networks.
\newblock {\em Phys. Rev. E}, 71:065103.

\bibitem[Pereira et~al., 2016]{Pereira.2016}
Pereira, J.~B., Mijalkov, M., Kakaei, E., Mecocci, P., Vellas, B., Tsolaki, M.,
  Kloszewska, I., Soininen, H., Spenger, C., Lovestone, S., Simmons, A.,
  Wahlund, L.-O., Volpe, G., and Westman, E. (2016).
\newblock Disrupted network topology in patients with stable and progressive
  mild cognitive impairment and alzheimer's disease.
\newblock {\em Cerebral Cortex}, 26(8):3476--3493.

\bibitem[Ravikumar et~al., 2011]{Ravikumar.2011}
Ravikumar, P., Wainwright, M.~J., Raskutti, G., and Yu, B. (2011).
\newblock High-dimensional covariance estimation by minimizing ℓ1-penalized
  log-determinant divergence.
\newblock {\em Electronic Journal of Statistics}, 5(0):935--980.

\bibitem[Rubinov and Sporns, 2010]{Rubinov.2010}
Rubinov, M. and Sporns, O. (2010).
\newblock Complex network measures of brain connectivity: Uses and
  interpretations.
\newblock {\em NeuroImage}, 52(3):1059--1069.

\bibitem[Ryali et~al., 2012]{Ryali.2012}
Ryali, S., Chen, T., Supekar, K., and Menon, V. (2012).
\newblock Estimation of functional connectivity in fmri data using stability
  selection-based sparse partial correlation with elastic net penalty.
\newblock {\em NeuroImage}, 59(4):3852--3861.

\bibitem[Sakr et~al., 2019]{Sakr.2019}
Sakr, F.~A., Grothe, M.~J., Cavedo, E., Jelistratova, I., Habert, M.-O., Dyrba,
  M., Gonzalez-Escamilla, G., Bertin, H., Locatelli, M., Lehericy, S., Teipel,
  S., Dubois, B., and Hampel, H. (2019).
\newblock {Applicability of in vivo staging of regional amyloid burden in a
  cognitively normal cohort with subjective memory complaints: the
  INSIGHT-preAD study}.
\newblock {\em Alzheimer's research {\&} therapy}, 11(1):15.

\bibitem[Savio et~al., 2017]{Savio.2017}
Savio, A., F{\"u}nger, S., Tahmasian, M., Rachakonda, S., Manoliu, A., Sorg,
  C., Grimmer, T., Calhoun, V., Drzezga, A., Riedl, V., and Yakushev, I.
  (2017).
\newblock Resting-state networks as simultaneously measured with functional mri
  and pet.
\newblock {\em Journal of nuclear medicine : official publication, Society of
  Nuclear Medicine}, 58(8):1314--1317.

\bibitem[Schultz et~al., 2017]{Schultz.2017}
Schultz, A.~P., Chhatwal, J.~P., Hedden, T., Mormino, E.~C., Hanseeuw, B.~J.,
  Sepulcre, J., Huijbers, W., LaPoint, M., Buckley, R.~F., Johnson, K.~A., and
  Sperling, R.~A. (2017).
\newblock Phases of hyperconnectivity and hypoconnectivity in the default mode
  and salience networks track with amyloid and tau in clinically normal
  individuals.
\newblock {\em Journal of Neuroscience}, 37(16):4323--4331.

\bibitem[Seeley et~al., 2009]{Seeley.2009}
Seeley, W.~W., Crawford, R.~K., Zhou, J., Miller, B.~L., and Greicius, M.~D.
  (2009).
\newblock Neurodegenerative diseases target large-scale human brain networks.
\newblock {\em Neuron}, 62(1):42--52.

\bibitem[Sepulcre et~al., 2013]{Sepulcre.2013}
Sepulcre, J., Sabuncu, M.~R., Becker, A., Sperling, R., and Johnson, K.~A.
  (2013).
\newblock In vivo characterization of the early states of the amyloid-beta
  network.
\newblock {\em Brain}, 136(Pt 7):2239--2252.

\bibitem[Sepulcre et~al., 2017]{Sepulcre.2017}
Sepulcre, J., Sabuncu, M.~R., Li, Q., {El Fakhri}, G., Sperling, R., and
  Johnson, K.~A. (2017).
\newblock Tau and amyloid beta proteins distinctively associate to functional
  network changes in the aging brain.
\newblock {\em Alzheimer's {\&} Dementia}, 13(11):1261--1269.

\bibitem[Spetsieris et~al., 2015]{Spetsieris.2015}
Spetsieris, P.~G., Ko, J.~H., Tang, C.~C., Nazem, A., Sako, W., Peng, S., Ma,
  Y., Dhawan, V., and Eidelberg, D. (2015).
\newblock Metabolic resting-state brain networks in health and disease.
\newblock {\em Proceedings of the National Academy of Sciences of the United
  States of America}, 112(8):2563--2568.

\bibitem[Stam et~al., 2006]{Stam.2006}
Stam, C., Jones, B., Nolte, G., Breakspear, M., and Scheltens, P. (2006).
\newblock Small-world networks and functional connectivity in alzheimer's
  disease.
\newblock {\em Cerebral Cortex}, 17(1):92--99.

\bibitem[Sun et~al., 2014]{Sun.2014}
Sun, S., Zhu, Y., and Xu, J. (2014).
\newblock { Adaptive Variable Clustering in Gaussian Graphical Models}.
\newblock In Kaski, S. and Corander, J., editors, {\em Proceedings of the
  Seventeenth International Conference on Artificial Intelligence and
  Statistics}, volume~33 of {\em Proceedings of Machine Learning Research},
  pages 931--939, Reykjavik, Iceland. PMLR.

\bibitem[Teipel et~al., 2015a]{Teipel.2015}
Teipel, S., Drzezga, A., Grothe, M.~J., Barthel, H., Ch{\'e}telat, G., Schuff,
  N., Skudlarski, P., Cavedo, E., Frisoni, G.~B., Hoffmann, W., Thyrian, J.~R.,
  Fox, C., Minoshima, S., Sabri, O., and Fellgiebel, A. (2015a).
\newblock Multimodal imaging in {A}lzheimer's disease: Validity and usefulness
  for early detection.
\newblock {\em The Lancet Neurology}, 14(10):1037--1053.

\bibitem[Teipel and Grothe, 2016]{Teipel.2016}
Teipel, S. and Grothe, M.~J. (2016).
\newblock Does posterior cingulate hypometabolism result from disconnection or
  local pathology across preclinical and clinical stages of {A}lzheimer's
  disease?
\newblock {\em European journal of nuclear medicine and molecular imaging},
  43(3):526--536.

\bibitem[Teipel et~al., 2016]{Teipel.2016b}
Teipel, S., Grothe, M.~J., Zhou, J., Sepulcre, J., Dyrba, M., Sorg, C., and
  Babiloni, C. (2016).
\newblock Measuring cortical connectivity in alzheimer's disease as a brain
  neural network pathology: Toward clinical applications.
\newblock {\em Journal of the International Neuropsychological Society},
  22(2):138--163.

\bibitem[Teipel et~al., 2015b]{Teipel.2015b}
Teipel, S.~J., Kurth, J., Krause, B., and Grothe, M.~J. (2015b).
\newblock The relative importance of imaging markers for the prediction of
  {A}lzheimer's disease dementia in mild cognitive impairment - beyond
  classical regression.
\newblock {\em NeuroImage. Clinical}, 8:583--593.

\bibitem[Tijms et~al., 2013]{Tijms.2013}
Tijms, B.~M., M{\"o}ller, C., Vrenken, H., Wink, A.~M., de~Haan, W., {van der
  Flier, Wiesje M}, Stam, C.~J., Scheltens, P., and Barkhof, F. (2013).
\newblock Single-subject grey matter graphs in alzheimer's disease.
\newblock {\em PLoS ONE}, 8(3):e58921.

\bibitem[Titov et~al., 2017]{Titov.2017}
Titov, D., Diehl-Schmid, J., Shi, K., Perneczky, R., Zou, N., Grimmer, T., Li,
  J., Drzezga, A., and Yakushev, I. (2017).
\newblock Metabolic connectivity for differential diagnosis of dementing
  disorders.
\newblock {\em Journal of Cerebral Blood Flow {\&} Metabolism}, 37(1):252--262.

\bibitem[Torok et~al., 2018]{Torok.2018}
Torok, J., Maia, P.~D., Powell, F., Pandya, S., and Raj, A. (2018).
\newblock A method for inferring regional origins of neurodegeneration.
\newblock {\em Brain}.

\bibitem[Villain et~al., 2010]{Villain.2010}
Villain, N., Fouquet, M., Baron, J.-C., M{\'e}zenge, F., Landeau, B., de~{La
  Sayette}, V., Viader, F., Eustache, F., Desgranges, B., and Ch{\'e}telat, G.
  (2010).
\newblock Sequential relationships between grey matter and white matter atrophy
  and brain metabolic abnormalities in early {A}lzheimer's disease.
\newblock {\em Brain}, 133(11):3301--3314.

\bibitem[Voevodskaya et~al., 2017]{Voevodskaya.2017}
Voevodskaya, O., Pereira, J.~B., Volpe, G., Lindberg, O., Stomrud, E., {van
  Westen}, D., Westman, E., and Hansson, O. (2017).
\newblock Altered structural network organization in cognitively normal
  individuals with amyloid pathology.
\newblock {\em Neurobiology of aging}, 64:15--24.

\bibitem[Wang et~al., 2016]{Wang.2016}
Wang, Y., Kang, J., Kemmer, P.~B., and Guo, Y. (2016).
\newblock An efficient and reliable statistical method for estimating
  functional connectivity in large scale brain networks using partial
  correlation.
\newblock {\em Frontiers in Neuroscience}, 10:123.

\bibitem[Watts and Strogatz, 1998]{Watts.1998}
Watts, D.~J. and Strogatz, S.~H. (1998).
\newblock Collective dynamics of 'small-world' networks.
\newblock {\em Nature}, 393(6684):440--442.

\bibitem[Weise et~al., 2018]{Weise.2018}
Weise, C.~M., Chen, K., Chen, Y., Kuang, X., Savage, C.~R., and Reiman, E.~M.
  (2018).
\newblock Left lateralized cerebral glucose metabolism declines in amyloid-beta
  positive persons with mild cognitive impairment.
\newblock {\em NeuroImage. Clinical}, 20:286--296.

\bibitem[Yao et~al., 2010]{Yao.2010}
Yao, Z., Zhang, Y., Lin, L., Zhou, Y., Xu, C., and Jiang, T. (2010).
\newblock Abnormal cortical networks in mild cognitive impairment and
  alzheimer's disease.
\newblock {\em PLoS Computational Biology}, 6(11):e1001006.

\bibitem[Zhou et~al., 2012]{Zhou.2012}
Zhou, J., Gennatas, E.~D., Kramer, J.~H., Miller, B.~L., and Seeley, W.~W.
  (2012).
\newblock Predicting regional neurodegeneration from the healthy brain
  functional connectome.
\newblock {\em Neuron}, 73(6):1216--1227.

\end{thebibliography}

\clearpage
\newpage
\begin{landscape}
\section{Tables}

\begin{table}[!h]
	\centering
		\caption{Sample characteristics.}
		\begin{tabular}{l r r r r}
		\hline
						        & {\bf CN}	& {\bf EMCI}		& {\bf LMCI}	& {\bf AD}	\\
								&				&				&		&\\
		\hline
		Sample size	(female) 	& $254 (130)$ 	&	$309 (135)$ &	$220 (93)$ &	$189 (80)$ \\
		Age (SD)				& $75.4 \pm 6.6$ &	$71.6 \pm 7.5 $ * &	$74.1 \pm 8.1$ &	$75.0 \pm 8.0$ \\
		Education (SD)	     	& $16.4 \pm 2.7$ &	$16.0 \pm 2.6$ &	$16.2 \pm 2.8$ &	$15.9 \pm 2.7$ \\
		MMSE (SD)				& $29.1 \pm 1.2$ &	$28.3 \pm 1.6$ * &	$27.6 \pm 1.9$ * &	$22.6 \pm 3.2$ * \\
		Delayed recall (SD)	 	& $7.6 \pm 4.1$ &	$5.7 \pm 4.0$ * &	$3.2 \pm 3.7$ *	&	$0.8 \pm 1.9$ * \\
		\hline
		\end{tabular}
		\begin{flushleft}Gender distribution did not differ significantly between groups ($P=0.15$, chi-square test). 
			Asterisks indicate significant difference between groups ($P<0.05$) based on pairwise two-sample t-test with CN as reference group.
			CN: cognitively healthy elderly controls, EMCI/LMCI: early and late amnestic mild cognitive impairment, AD: Alzheimer's dementia, MMSE: Mini-Mental State Examination, delayed recall: number of remembered words out of a 15-item wordlist of the Rey Auditory Verbal Learning Test.
		\end{flushleft}
	\label{tab:sample}
\end{table}

\begin{table}[!h]
  \centering
	\caption{P-values for the group comparison of partial correlation graph statistics (Figure~\ref{fig:graph_statistics}).}
	\begin{tabular}{llrrrrrrrrr}
    \hline
		&	& \multicolumn{3}{c}{Amyloid-$\beta$} 	& \multicolumn{3}{c}{Metabolism} 	& \multicolumn{3}{c}{Volume} \\
		&       & EMCI  	& LMCI  	& AD    	& EMCI  	& LMCI  	& AD    	& EMCI  	& LMCI  	& AD \\
    \hline
		Clustering coefficient
		& CN 	& 0.167		& 0.999 	& 0.178		& 0.323 	& 0.021		& 0.718 	& 0.009		& 0.977		& 0.999 \\
		& EMCI  &       	& 0.183 	& $<0.001$ 	&       	& 0.630 	& 0.031		&       	& 0.030 	& 0.012 \\
		& LMCI  &      		&       	& 0.162 	&       	&       	& $<0.001$ 	&       	&       	& 0.990 \\
		Path length
		& CN 	& 0.264		& $<0.001$ 	& 0.630		& 0.015 	& 0.001		& 0.357 	& 0.106		& 0.664		& 0.005 \\
		& EMCI  &       	& 0.189 	& 0.922	 	&       	& 0.884	 	& $<0.001$	&       	& 0.667 	& $<0.001$ \\
		& LMCI  &      		&       	& 0.044 	&       	&       	& $<0.001$ 	&       	&       	& $<0.001$ \\
		Small-world coefficient
		& CN 	& 0.101		& 0.940	 	& 0.301		& 0.184 	& 0.002		& 0.701 	& 0.011		& 0.967 	& 0.987 \\
		& EMCI  &       	& 0.313 	& $<0.001$ 	&       	& 0.411 	& 0.011		&       	& 0.042 	& 0.029 \\
		& LMCI  &      		&       	& 0.096 	&       	&       	& $<0.001$ 	&       	&       	& 0.999 \\
    \hline
	\end{tabular}%
	\begin{flushleft}Adjusted P-values from Tukey's honest significant difference tests, controlling for family-wise error rate within each comparison block.
	CN: cognitively healthy elderly controls, EMCI/LMCI: early and late amnestic mild cognitive impairment, AD: Alzheimer's dementia.\end{flushleft}
  \label{tab:pvals}%
\end{table}%

\end{landscape}

\clearpage
\newpage
\section*{Figure legends}

\textbf{Figure 1.} Simple example for spurious correlations.
	(A) True dependency graph. The node $u$ is statistically independent from $v$ given the node $dis$, formally $p(u,v|dis) = p(u|dis)p(v|dis)$.
	(B) Pearson correlation matrix, showing a "spurious" correlation between nodes u and v. 
		Notably, when considering only $u$ and $v$ alone, the independence assumption does not hold; formally $p(u,v) \neq p(u)p(v)$.
	(C) Partial correlation matrix derived from Gaussian graphical models. 
		Using this model, we can approximately recover the underlying dependency structure, with $u \perp v | dis \implies cor(u,v|dis)=0$.

\textbf{Figure 2.} Pearson correlation matrix (left) and partial correlation matrix (right) for the three imaging modalities (left hemisphere data only) estimated for the combined data of EMCI, LMCI and AD patients.
	For better readability, each individual block of the partial correlation matrix is shown in Figures~\ref{fig:parcor-mat-lh-amy-amy}--\ref{fig:parcor-mat-lh-vol-vol} and Supplementary Figures \ref{sfig:parcor-mat-lh-amy-fdg}--\ref{sfig:parcor-mat-lh-amy-vol}.\\
	EMCI/LMCI: early and late amnestic mild cognitive impairment, AD: Alzheimer's dementia.

\textbf{Figure 3.} Partial correlation matrix for glucose amyloid-$\beta$ in the left hemisphere estimated for the combined data of EMCI, LMCI and AD patients. Averaged over ten repetitions. Associations of lowest magnitude were not present in all iterations.\\
EMCI/LMCI: early and late amnestic mild cognitive impairment, AD: Alzheimer's dementia, amy: amyloid-$\beta$.

\textbf{Figure 4.} Partial correlation matrix for glucose metabolism in the left hemisphere estimated for the combined data of EMCI, LMCI and AD patients. Averaged over ten repetitions. Associations of lowest magnitude were not present in all iterations.\\
EMCI/LMCI: early and late amnestic mild cognitive impairment, AD: Alzheimer's dementia, metab: glucose metabolism.

\textbf{Figure 5.} Partial correlation matrix for gray matter volume in the left hemisphere estimated for the combined data of EMCI, LMCI and AD patients. Averaged over ten repetitions. Associations of lowest magnitude were not present in all iterations.\\
EMCI/LMCI: early and late amnestic mild cognitive impairment, AD: Alzheimer's dementia, vol: gray matter volume.

\textbf{Figure 6.} Partial correlation matrix for amyloid-$\beta$ in the left hemisphere by group. Averaged over ten repetitions. Associations of lowest magnitude were not present in all iterations.\\
CN: cognitively healthy elderly controls, EMCI/LMCI: early and late amnestic mild cognitive impairment, AD: Alzheimer's dementia, amy: amyloid-$\beta$.

\textbf{Figure 7.} Partial correlation matrix for glucose metabolism in the left hemisphere by group. Averaged over ten repetitions. Associations of lowest magnitude were not present in all iterations.\\
CN: cognitively healthy elderly controls, EMCI/LMCI: early and late amnestic mild cognitive impairment, AD: Alzheimer's dementia, metab: glucose metabolism.

\textbf{Figure 8.} Partial correlation matrix for gray matter volume in the left hemisphere by group. Averaged over ten repetitions. Associations of lowest magnitude were not present in all iterations.\\
CN: cognitively healthy elderly controls, EMCI/LMCI: early and late amnestic mild cognitive impairment, AD: Alzheimer's dementia, vol: gray matter volume.

\textbf{Figure 9.} Comparison of graph statistics for the partial correlation matrices of the left hemisphere stratified by diagnostic group and image modality.
	Estimates based on Gaussian graphical models using multimodal neuroimaging data.
	The distribution of the weighted clustering coefficient, characteristic weighted path length, and small-world coefficient for individual brain regions is shown. 
	\changed{Boxes display median, first and third quartile of the distributions, and whiskers indicate $\pm1.5 \times$interquartile range.
	All blocks showed significant differences in mean between groups, one-way analysis of variance (ANOVA), $df=215$, $F\geq4$, $p<0.01$, $\eta^2 \geq 0.055$. Details are given in Supplementary Tab.~\ref{stab:anova}.
	P-values for Tukey's honest significant difference tests are given in Tab.~\ref{tab:pvals}.}\\
	CN: cognitively healthy elderly controls, EMCI/LMCI: early and late amnestic mild cognitive impairment, AD: Alzheimer's dementia, amy: amyloid-$\beta$, metab: glucose metabolism, vol: gray matter volume.

\clearpage
\newpage
\section{Figures}

\clearpage
\renewcommand\thefigure{S\arabic{figure}}
\setcounter{figure}{0}
\renewcommand\thetable{S\arabic{table}}
\setcounter{table}{0}

\newpage
\section{Supplementary material}


\begin{figure*}[!ht]
	\centering
		\includegraphics[width=\textwidth]{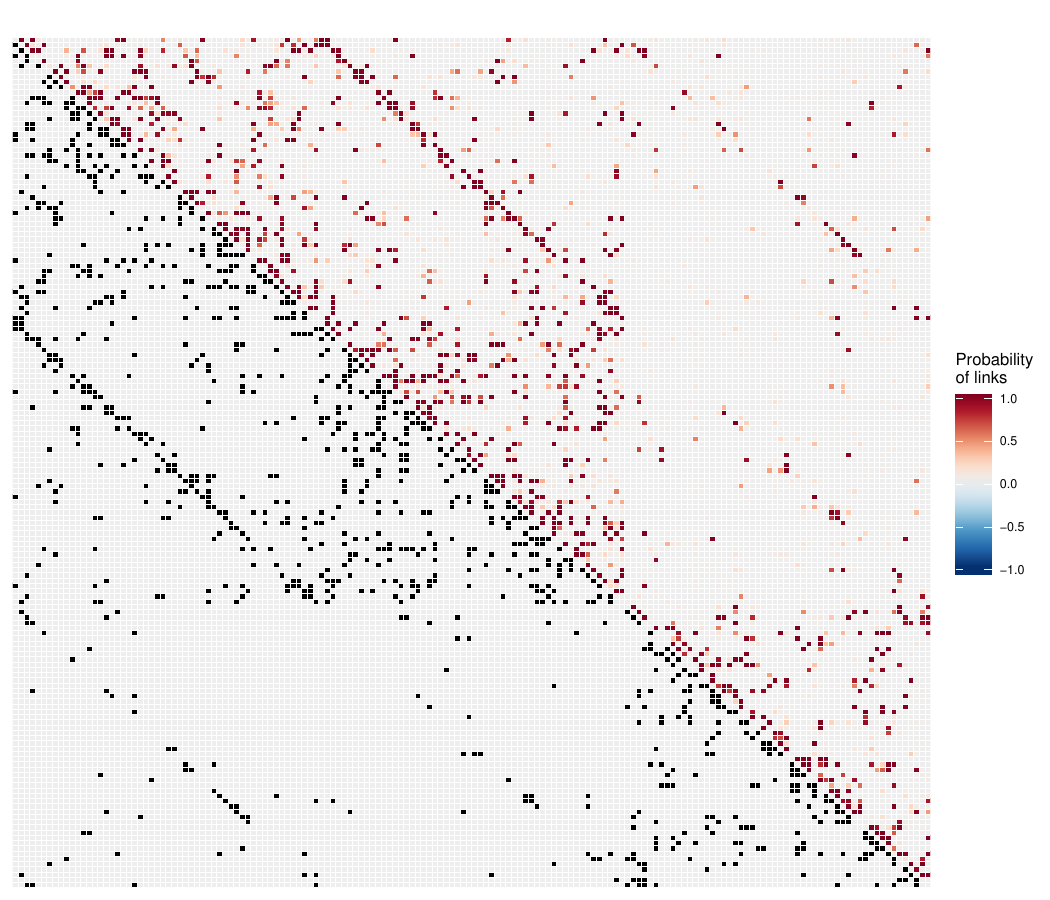}
	\caption{Probability of estimated edges for the left hemisphere. The upper right part provides the raw probability of each edge to exist. The lower left part indicates the selected edges exceeding the threshold of $P_{avg}>0.5$.}
	\label{sfig:plinks-lh}
\end{figure*}

\begin{figure*}[!ht]
	\centering
		\includegraphics[width=\textwidth]{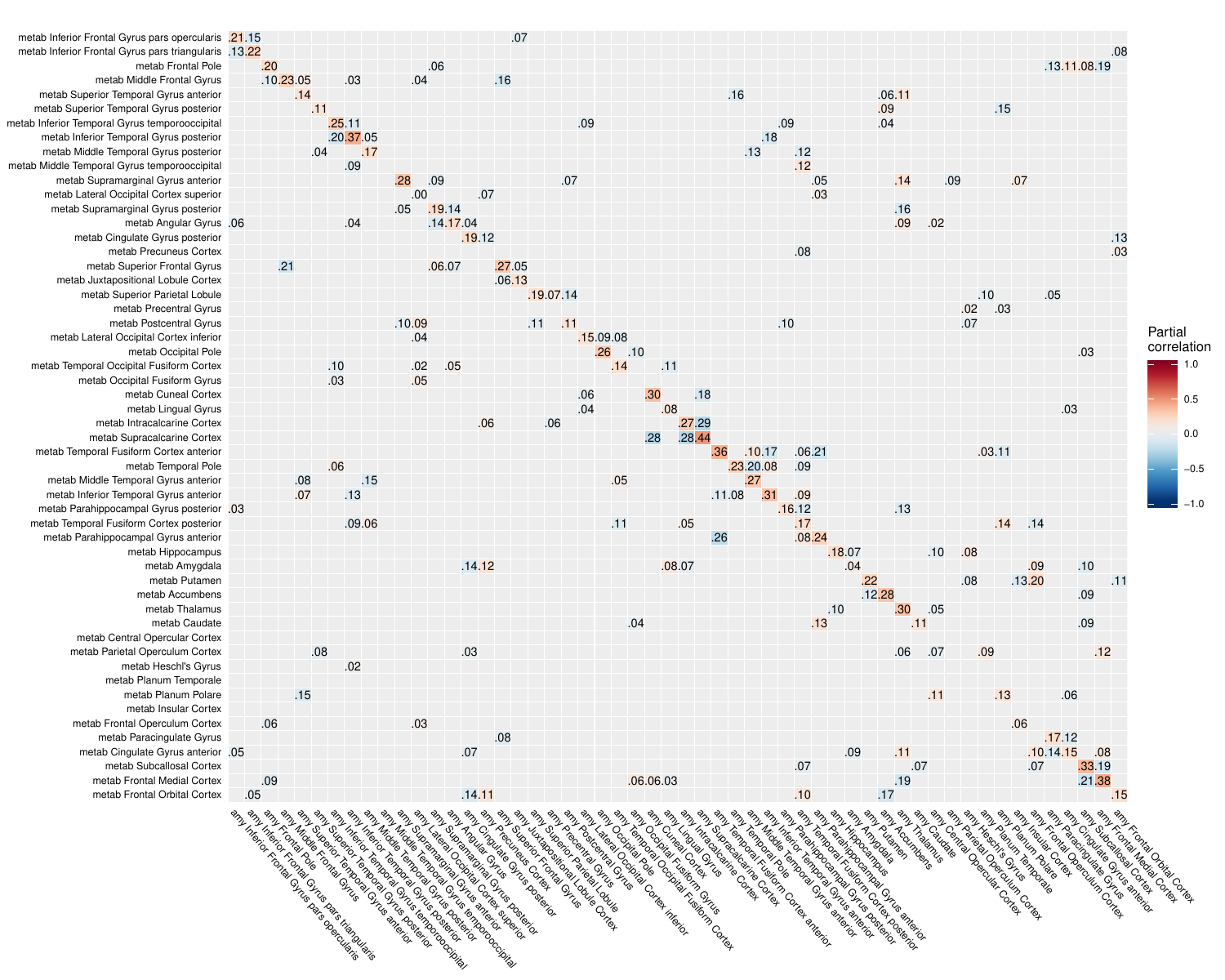}
	\caption{Partial correlation matrix for amyloid-$\beta$ deposition and glucose metabolism in the left hemisphere estimated for the combined data of EMCI, LMCI and AD patients. Averaged over ten repetitions. Associations of lowest magnitude were not present in all iterations.\\
EMCI/LMCI: early and late amnestic mild cognitive impairment, AD: Alzheimer's dementia, amy: amyloid-$\beta$, metab: glucose metabolism.}
	\label{sfig:parcor-mat-lh-amy-fdg}
\end{figure*}

\begin{figure*}[!ht]
	\centering
		\includegraphics[width=\textwidth]{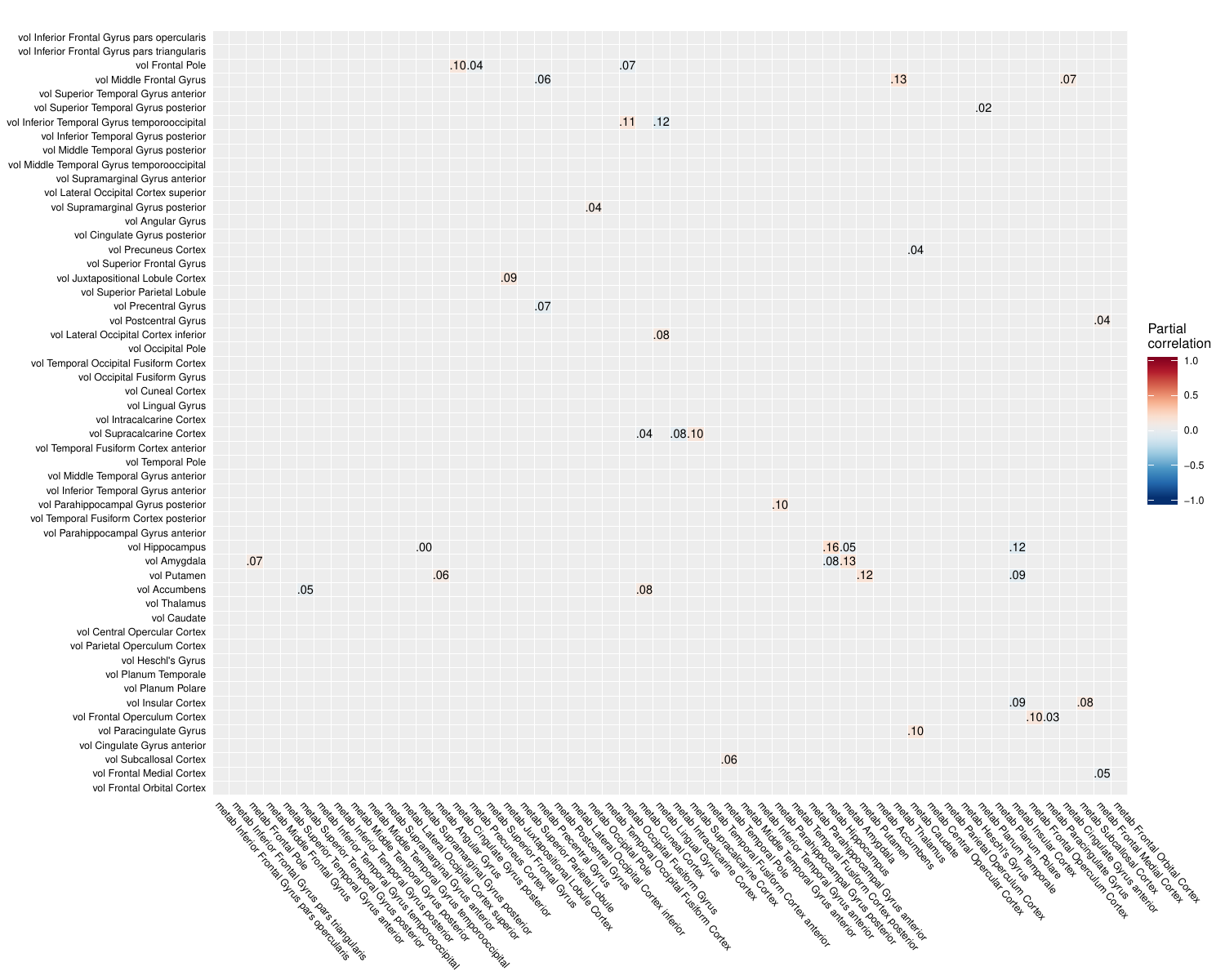}
	\caption{Partial correlation matrix for glucose metabolism and gray matter volume in the left hemisphere estimated for the combined data of EMCI, LMCI and AD patients. Averaged over ten repetitions. Associations of lowest magnitude were not present in all iterations.\\
EMCI/LMCI: early and late amnestic mild cognitive impairment, AD: Alzheimer's dementia, metab: glucose metabolism, vol: gray matter volume.}
	\label{sfig:parcor-mat-lh-fdg-vol}
\end{figure*}

\begin{figure*}[!ht]
	\centering
		\includegraphics[width=\textwidth]{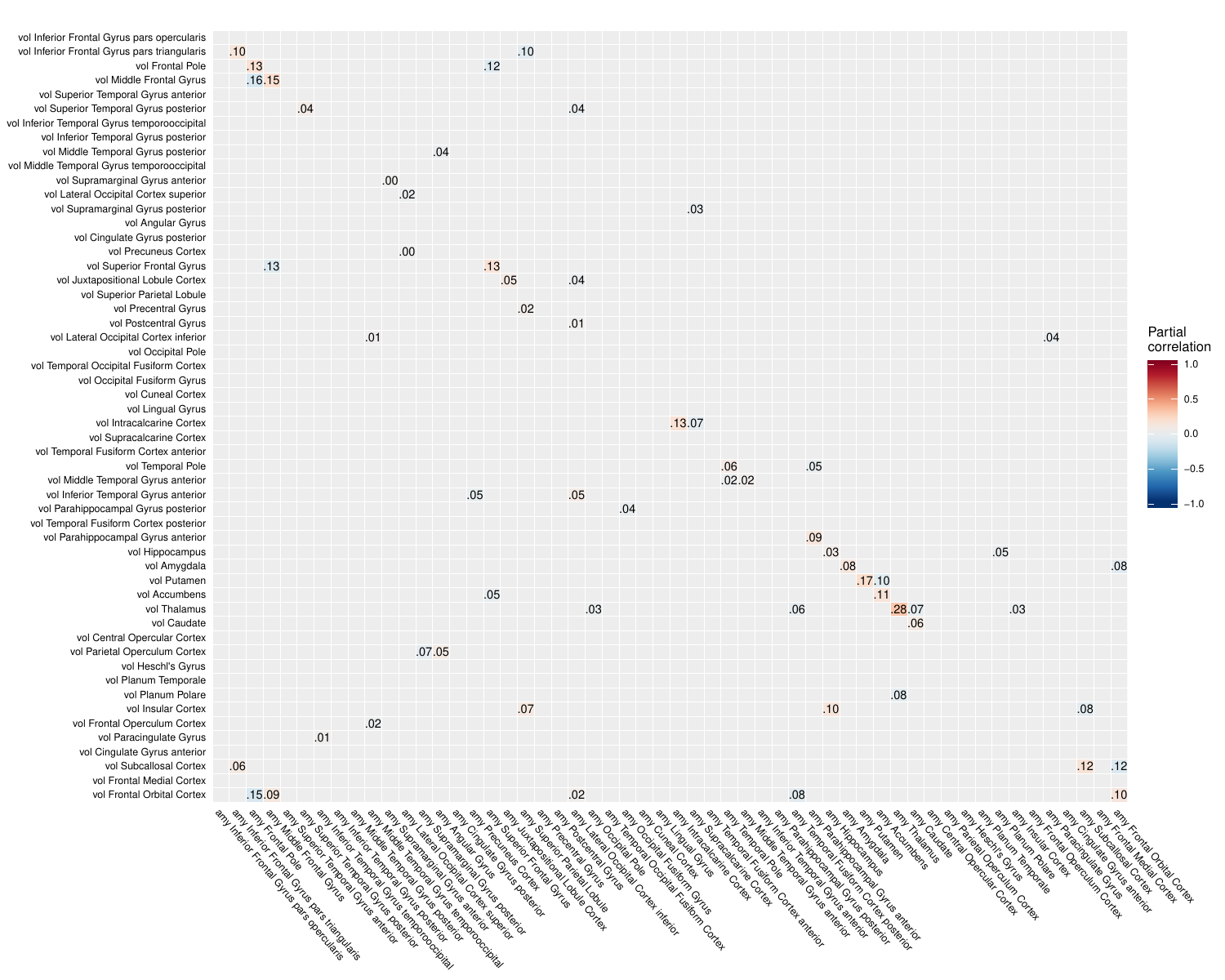}
	\caption{Partial correlation matrix for amyloid-$\beta$ deposition and gray matter volume in the left hemisphere estimated for the combined data of EMCI, LMCI and AD patients. Averaged over ten repetitions. Associations of lowest magnitude were not present in all iterations.\\
EMCI/LMCI: early and late amnestic mild cognitive impairment, AD: Alzheimer's dementia, amy: amyloid-$\beta$, vol: gray matter volume.}
	\label{sfig:parcor-mat-lh-amy-vol}
\end{figure*}

\begin{figure*}[!ht]
	\centering
		\includegraphics[width=0.8\textwidth]{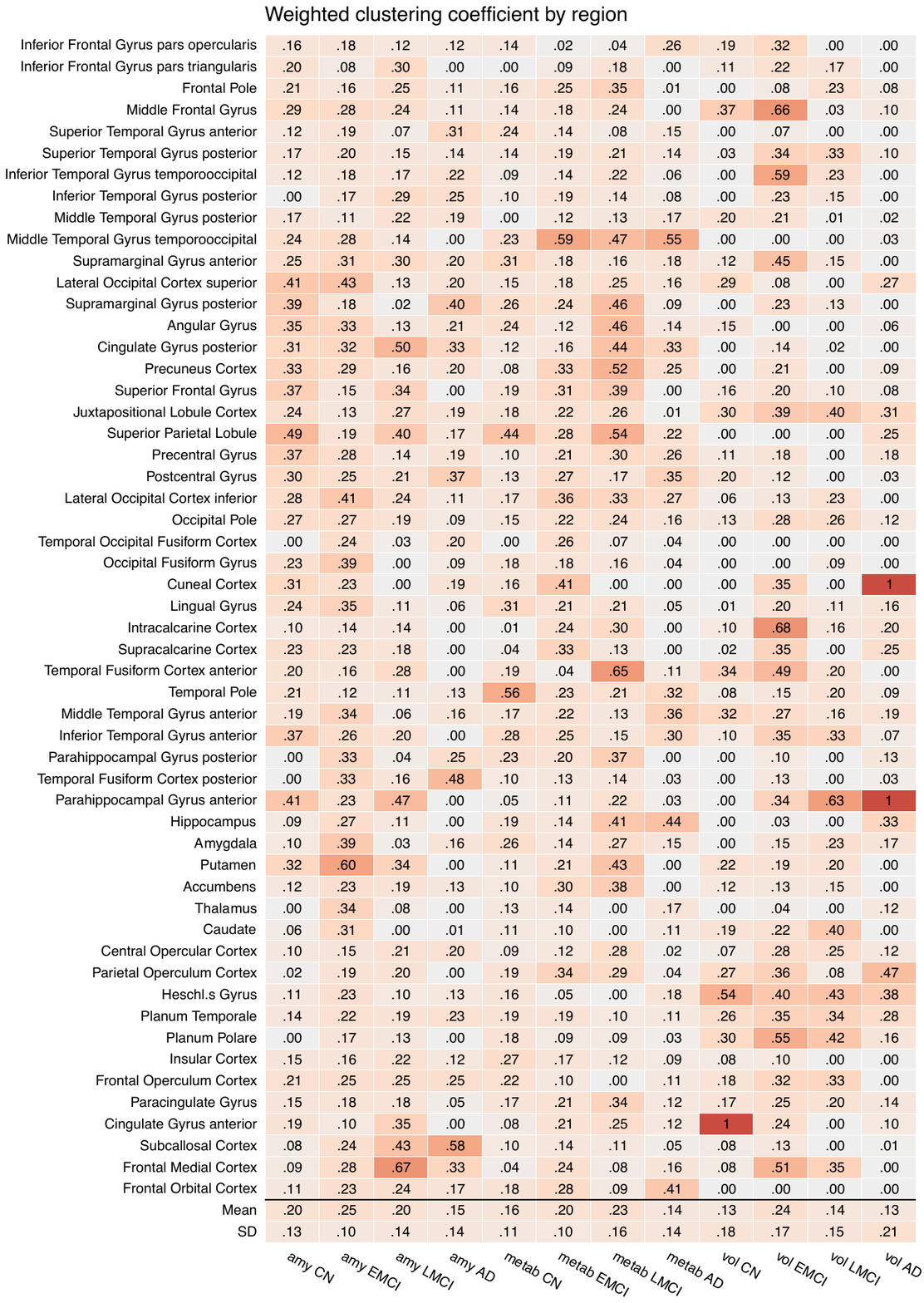}
	\caption{Comparison of weighted clustering coefficient stratified by brain region, diagnostic group and modality for the partial correlation matrices of the left hemisphere. Averaged over ten repetitions.\\
CN: cognitively healthy elderly controls, EMCI/LMCI: early and late amnestic mild cognitive impairment, AD: Alzheimer's dementia, amy: amyloid-$\beta$, metab: glucose metabolism, vol: gray matter volume.}
	\label{sfig:cc_by_region}
\end{figure*}

\begin{figure*}[!ht]
	\centering
		\includegraphics[width=0.8\textwidth]{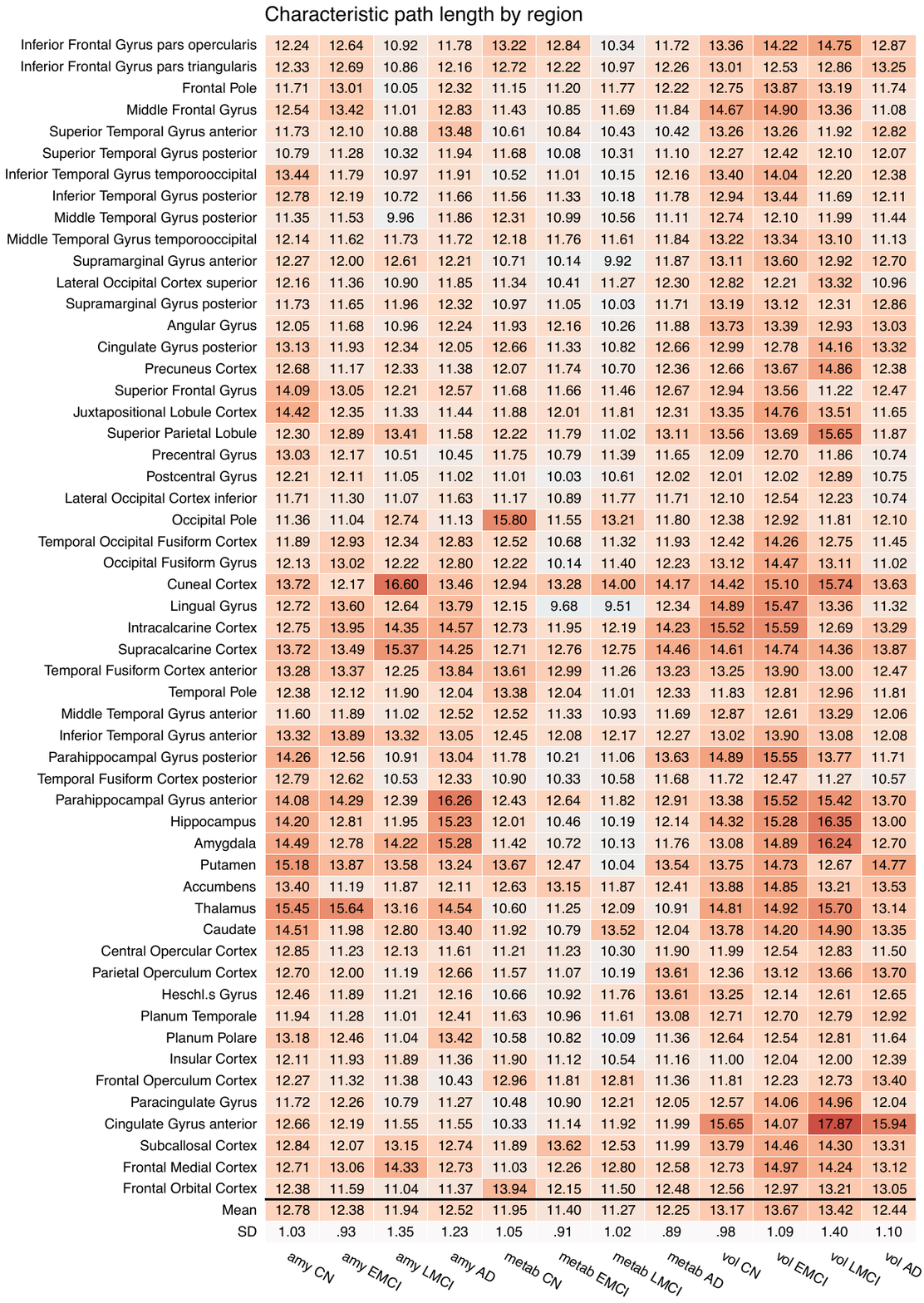}
	\caption{Comparison of characteristic path length stratified by brain region, diagnostic group and modality for the partial correlation matrices of the left hemisphere. Averaged over ten repetitions.\\
CN: cognitively healthy elderly controls, EMCI/LMCI: early and late amnestic mild cognitive impairment, AD: Alzheimer's dementia, amy: amyloid-$\beta$, metab: glucose metabolism, vol: gray matter volume.}
	\label{sfig:pl_by_region}
\end{figure*}

\begin{figure*}[!ht]
	\centering
		\includegraphics[width=0.8\textwidth]{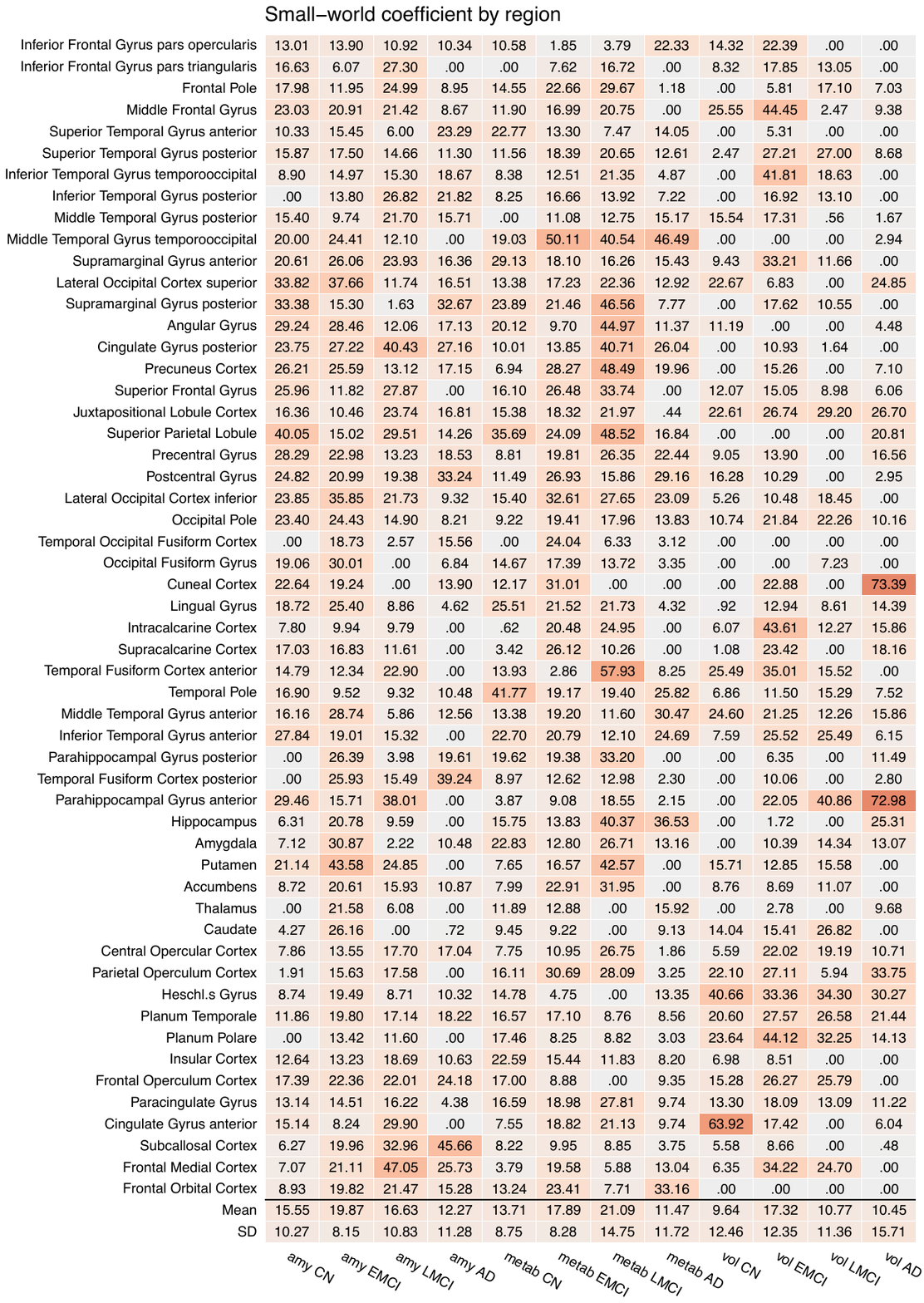}
	\caption{Comparison of small-world coefficient stratified by brain region, diagnostic group and modality for the partial correlation matrices of the left hemisphere. For better readability, individual values were upscaled by a factor of 1,000. Averaged over ten repetitions.\\
CN: cognitively healthy elderly controls, EMCI/LMCI: early and late amnestic mild cognitive impairment, AD: Alzheimer's dementia, amy: amyloid-$\beta$, metab: glucose metabolism, vol: gray matter volume.}
	\label{sfig:swc_by_region}
\end{figure*}

\begin{figure*}[!ht]
	\centering
		\includegraphics[width=0.9\textwidth]{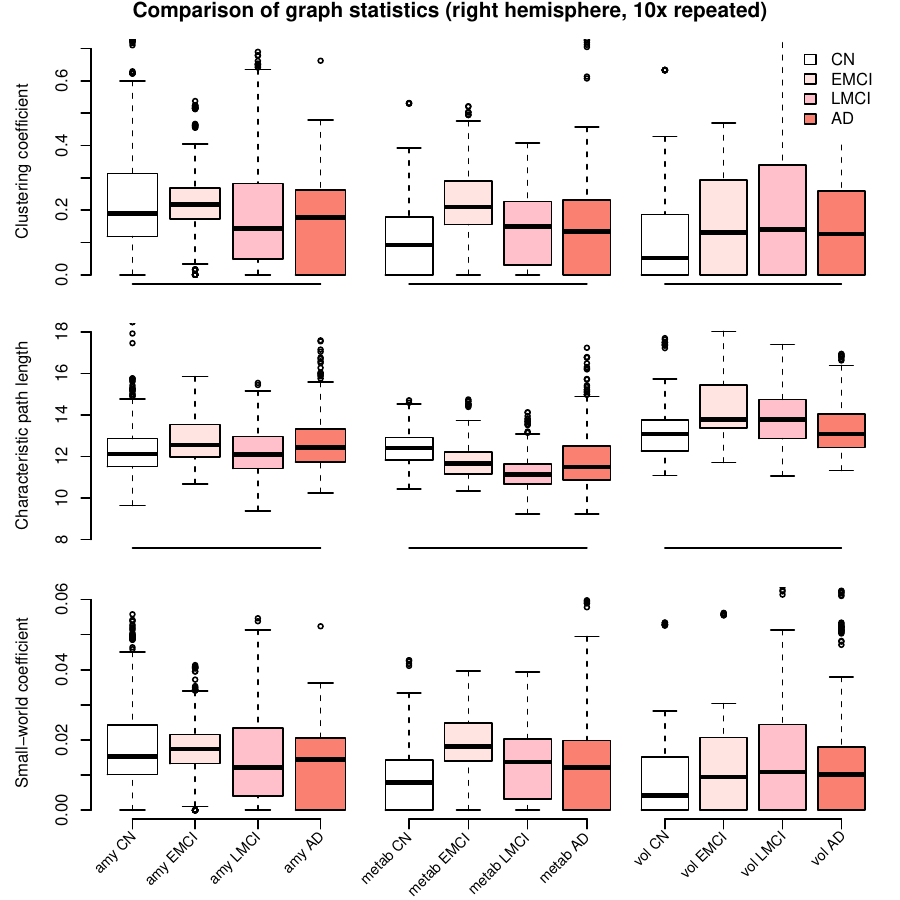}
	\caption{Comparison of graph statistics for the partial correlation matrices of the \textbf{right} hemisphere stratified by diagnostic group and image modality.
	Estimates based on Gaussian graphical models using multimodal neuroimaging data.
	The distribution of the weighted clustering coefficient, characteristic weighted path length, and small-world coefficient for individual brain regions is shown.
	Boxes display median, first and third quartile of the distributions, and whiskers indicate $\pm1.5 \times$interquartile range.\\
	CN: cognitively healthy elderly controls, EMCI/LMCI: early and late amnestic mild cognitive impairment, AD: Alzheimer's dementia, amy: amyloid-$\beta$, metab: glucose metabolism, vol: gray matter volume.}
	\label{sfig:graph_statistics_righthemi}
\end{figure*}

\begin{figure*}[!ht]
	\centering
		\includegraphics[width=\textwidth]{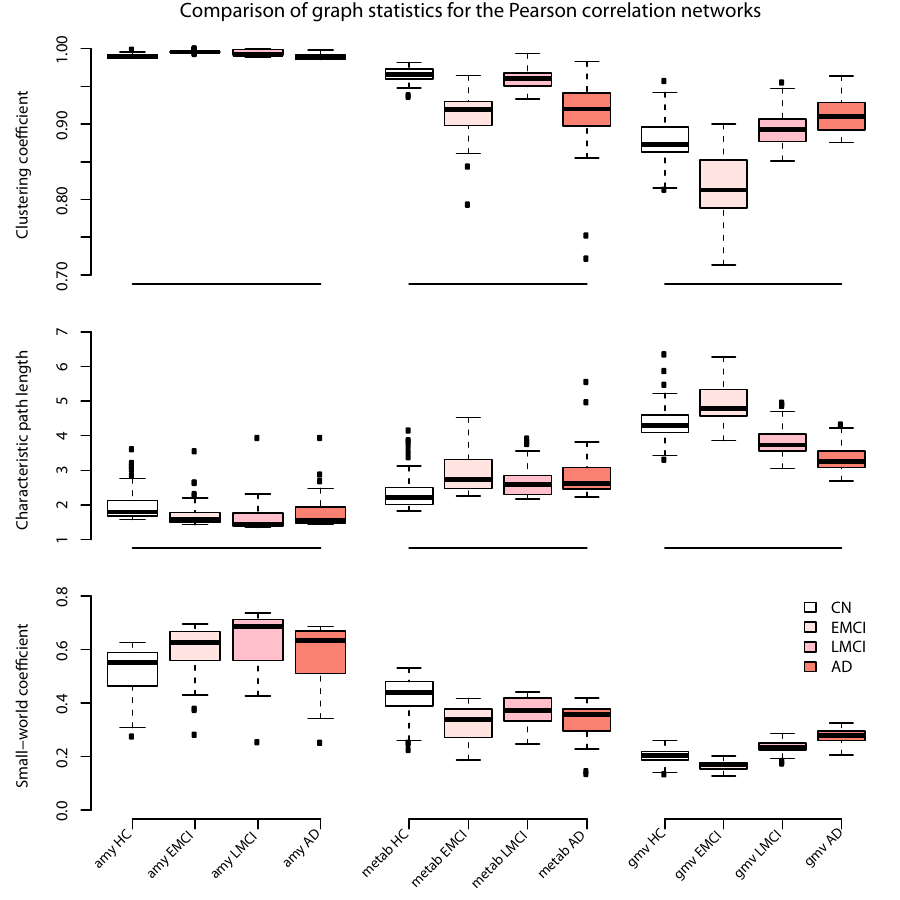}
	\caption{Comparison of graph statistics for the Pearson correlation matrices of the left hemisphere stratified by diagnostic group and image modality.
	The distribution of the weighted clustering coefficient, characteristic weighted path length, and small-world coefficient for individual brain regions is shown.
	Boxes display median, first and third quartile of the distributions, and whiskers indicate $\pm1.5 \times$interquartile range.
	Prior to calculating the graph measures, the correlation matrices were thresholded such that correlations with $p>0.05$, i.e. approximately $r<0.12$, were set to zero.\\
	CN: cognitively healthy elderly controls, EMCI/LMCI: early and late amnestic mild cognitive impairment, AD: Alzheimer's dementia, amy: amyloid-$\beta$, metab: glucose metabolism, vol: gray matter volume.}
	\label{sfig:graph_stats_Pearson}
\end{figure*}


\clearpage
\newpage

\begin{landscape}

\begin{table}[!ht]
	\centering
	\caption{P-values for the comparison of graph statistics based on Pearson correlation (Figure~\ref{sfig:graph_stats_Pearson}).}
	\begin{tabular}{llrrrrrrrrr}
		\hline
		&	& \multicolumn{3}{c}{Amyloid-$\beta$} 	& \multicolumn{3}{c}{Metabolism} 	& \multicolumn{3}{c}{Volume} \\
		&       & EMCI  	& LMCI  	& AD    	& EMCI  	& LMCI  	& AD    	& EMCI  	& LMCI  	& AD \\
		\hline
		Clustering coefficient
		& CN 	& $<0.001$	& $<0.001$ 	& 0.575		& $<0.001$ 	& 0.759 	& $<0.001$ 	& $<0.001$	& 0.103		& $<0.001$ \\
		& EMCI  &       	& $<0.001$ 	& $<0.001$ 	&       	& $<0.001$ 	& 0.973		&       	& $<0.001$ 	& $<0.001$ \\
		& LMCI  &      		&       	& $<0.001$ 	&       	&       	& $<0.001$ 	&       	&       	& 0.020	\\
		Path length
		& CN 	& 0.013		& $<0.001$ 	& 0.026		& $<0.001$ 	& 0.051		& $<0.001$ 	& $<0.001$	& $<0.001$	& $<0.001$ \\
		& EMCI  &       	& 0.437		& 0.996	 	&       	& 0.113	 	& 0.857		&       	& $<0.001$ 	& $<0.001$ \\
		& LMCI  &      		&       	& 0.315	 	&       	&       	& 0.464	 	&       	&       	& $<0.001$ \\
		Small-world coefficient
		& CN 	& $<0.001$	& $<0.001$ 	& 0.001		& $<0.001$ 	& $<0.001$	& $<0.001$ 	& $<0.001$	& $<0.001$	& $<0.001$ \\
		& EMCI  &       	& 0.085		& 0.991	 	&       	& 0.004 	& 0.797 	&       	& $<0.001$ 	& $<0.001$ \\
		& LMCI  &      		&       	& 0.040		&       	&       	& 0.055	 	&       	&       	& $<0.001$ \\
		\hline
	\end{tabular}%
	\begin{flushleft}Adjusted P-values from Tukey's honest significant difference tests, controlling for family-wise error rate within each comparison block.
	CN: cognitively normal controls, EMCI/LMCI: early/late amnestic mild cognitive impairment, AD: Alzheimer's dementia.\end{flushleft}
	\label{stab:pvals2}%
\end{table}%

\begin{table}[!h]
	\centering
		\caption{Analysis of variance (ANOVA) results for the graph statistics for the partial correlation networks in Figure~\ref{fig:graph_statistics}.}
		\begin{tabular}{l r r r r}
		\hline
						      		& 		& F-statistic	& P-value	& Effect size $\eta^2$	\\
		\hline
		Clustering coefficient 		& amy 	&	5.6			&	0.001		& 0.07				\\
		 							& metab &	6.2			&	$<0.001$	& 0.08				\\
		 		 					& vol 	&	4.6			&	0.004		& 0.06				\\
		Characteristic path length 	& amy 	&	5.1			&	0.002		& 0.07				\\
		 							& metab &	12.8		&	$<0.001$	& 0.15				\\
		 		 					& vol 	&	11.8		&	$<0.001$	& 0.14				\\
		Small-world coefficient		& amy 	&	5.6			&	0.001		& 0.07				\\
		 							& metab &	8.6			&	$<0.001$	& 0.11				\\
		 		 					& vol 	&	4.1			&	0.007		& 0.06				\\
		\hline
		\end{tabular}
		\begin{flushleft}df=215 for all models, amy: amyloid-$\beta$, metab: glucose metabolism, vol: gray matter volume.
		\end{flushleft}
	\label{stab:anova}
\end{table}

\end{landscape}

\end{document}